\documentclass[a4paper,12pt]{article}

\usepackage{geometry,slashed}
\usepackage
{
   amsmath,amssymb,graphicx,tabularx
}
\usepackage{soul}
\usepackage[usenames,dvipsnames]{xcolor} 
\usepackage{color}
\usepackage{hyperref}
\usepackage{caption,subcaption}
\usepackage{authblk}
\usepackage{epstopdf}
\usepackage{enumerate}
\usepackage{multirow}
\usepackage{siunitx}
\usepackage{float}
\usepackage{booktabs}
\usepackage{cite}
\usepackage{pdflscape} 
\usepackage{listings}
\usepackage{diagbox} 
\DeclareMathAlphabet{\pazocal}{OMS}{zplm}{m}{n}

\numberwithin{equation}{section}

\title{On improving NLO merging for $t \bar t W$ production}

\author{Rikkert Frederix\thanks{rikkert.frederix@thep.lu.se} }
\author{Ioannis Tsinikos\thanks{ioannis.tsinikos@thep.lu.se} }

\affil{\small Theoretical Particle Physics, Department of Astronomy and Theoretical Physics, Lund University, S\"olvegatan 14A, SE-223 62 Lund, Sweden}

\begin{document}

\maketitle

\vspace*{-9cm}
\noindent
{\small LU-TP 21-34}
\vspace*{9cm}

\begin{abstract}
We introduce an improvement to the {\sc\small FxFx} matrix element
merging procedure for $pp\to t \bar t W$ production at NLO in QCD with
one and/or two additional jets. The main modification is an improved
treatment of jets that are not logarithmically enhanced in the low
transverse-momentum regime. We provide predictions for the inclusive
cross section and the $t \bar t W$ differential distributions
including parton-shower effects. Taking also the NLO EW corrections
into account, this results in the most-accurate predictions for this
process to date. We further proceed to include the on-shell LO decays
of the $t \bar t W$ including the tree-level spin correlations within
the narrow-width approximation, focusing on the multi-lepton
signatures studied at the LHC. We find a $\sim\!\!30\%$ increase over
the NLO QCD prediction and large non-flat $K$-factors to differential
distributions.
\end{abstract}

\tableofcontents

\newpage

\section{Introduction}
\label{sec:intro}

With the results from the latest LHC run, both the ATLAS and CMS
collaborations have investigated the associated top-quark pair
production in association with a massive Weak vector boson. Both $t
\bar t Z$ and $t \bar t W$ production modes are measured at the
inclusive level~\cite{Aaboud:2019njj,Sirunyan:2017uzs} at 13
TeV. Recently, the experimental ATLAS and CMS groups have included
differential measurements and presented comparisons with theoretical
predictions for $t \bar t Z$~\cite{Aad:2021tya,CMS:2019too}. In both
these processes there is agreement between theory predictions and data
at the inclusive level, with a a slightly higher measured cross
section than predicted for $t \bar t W$. Regarding the $t \bar t Z$
production, this agreement is true also at differential level.

On top of these direct-measurement analyses, these processes are
dominant backgrounds to $t \bar t H$ and $t\bar t t \bar t$
production~\cite{ATLAS:2019nvo,CMS:2017vru,Aad:2020klt}. The
multi-lepton signatures of $t \bar t H$ and $t\bar t t \bar t$
production require a precise theoretical modeling of the backgrounds
in order to remove them. In these analyses a significant tension is
reported for $t \bar t W$, being the main irreducible background. In
the $t \bar t H$ multi-lepton analysis~\cite{ATLAS:2019nvo}, the
reported tension on $t \bar t W$ in the low jet multiplicity regime of
the two same sign lepton signal region results to a normalisation
factor of $1.56 \pm 0.3$ which reaches the value $1.68\pm 0.3$ in the
three-lepton signal region. Similarly, in the $t \bar t t \bar t$
multi-lepton analysis~\cite{Aad:2020klt}, the $t \bar t W + {\rm
  jets}$ validation region shows a higher than $\sim\!\!2$ ratio of data
over prediction in the high jet multiplicity regime, leading to a
normalisation factor of $1.6 \pm 0.3$ for this background. Both these
observations introduce large systematic uncertainties in the
experimental analyses and most importantly they indicate that in the
multi-lepton signal regions the theoretical prediction of the $t \bar
t W$ production is lower than the measurements.

The $t \bar t W$ process is studied theoretically in detail beyond NLO
in QCD at the production level, which corresponds to the
parton-unfolded level of the experimental analyses. Regarding the EW
corrections, the complete-NLO calculation~\cite{Frederix:2017wme} has
shown that the NLO EW ($\alpha_S^2\alpha^2$) corrections reduce the LO
cross section by $\sim\!\!-4 \%$ and the dominant sub-leading EW
corrections ($\alpha_S\alpha^3$) increase it by $\sim\!\!12 \%$. The
source of the surprisingly large contributions of order $\mathcal
O(\alpha_S\alpha^3)$ is identified to be the emerging $tW \rightarrow
tW$ scattering diagrams~\cite{Dror:2015nkp}. In
Ref.~\cite{Frederix:2020jzp} these contributions are studied with a
focus on their effect to the jet-multiplicity distributions. The most
accurate calculations to date for the $t \bar t W$ production have
matched the complete-NLO fixed-order calculation to threshold
resummation at NNLL
accuracy~\cite{Broggio:2019ewu,Kulesza:2020nfh}. In
Refs.~\cite{Broggio:2019ewu,Kulesza:2020nfh} the $t \bar t W$
production is calculated along with the $t \bar t Z$ and $t \bar t H$
processes. These works reveal a well-known feature of $t\bar{t}W$
production clearly. This final state is exclusively produced by
$q\bar{q}$ annihilation at LO, while for the other two processes the
$gg$ initial states also contribute at this order. On top of this, the
corrections coming from hard, non-logarithmically enhanced radiation are
large~\cite{Maltoni:2015ena,Frederix:2017wme}, resulting in a strong
dependence on the renormalisation scale; in particular, contrary to $t
\bar t Z$ and $t \bar t H$ production, for the $t \bar t W$ process
various choices for central values give significantly different
results, even at NNLL accuracy, as discussed in detail in
Refs.~\cite{Broggio:2019ewu} and \cite{Kulesza:2020nfh}. The absence
of the $gg$ contribution up to NNLO explains another feature of the $t
\bar t W$ process, that is the large central-peripheral $t\bar t$
asymmetry~\cite{Maltoni:2014zpa,Maltoni:2015ena,Broggio:2019ewu}.

By decaying the $t \bar t W$ resonances, one enters the multi-lepton and multi-jet signal regions (decay level) that are actually observed as signatures in the detectors. There are various advantages of performing the calculation at the decay level; one can apply selections and cuts to the final observables reproducing the fiducial region, calculate NLO corrections to the decays of the resonances, include off-shell effects along with non-resonant contributions and match the calculation to the parton shower in order to predict realistic jet-related observables. The caveat of this level of calculation is the increasing complexity of the Feynman-diagram structures that prevents one to study all these features simultaneously. At fixed order, the $t \bar t W$ in the three-lepton decay channel is calculated at NLO in QCD including off-shell effects and non-resonant top-quark contributions~\cite{Bevilacqua:2020pzy,Denner:2020hgg}. These calculations are followed by the complete-NLO calculation of the three-lepton final signature~\cite{Denner:2021hqi}. These works have demonstrated in detail the effects of the NLO QCD and EW corrections to the three-lepton decay as well as the off-shell effects to the final lepton and jet dependent differential distributions. The leptonic and $b-$jet asymmetries, calculated initially within the narrow-width approximation and LO decays~\cite{Maltoni:2014zpa}, are updated into including NLO QCD decays and off-shell effects~\cite{Bevilacqua:2020srb}.

The state of the art calculations described in the previous paragraph
do include NLO corrections to the decays and off-shell effects, but
they are fixed order calculations. The inclusion of the parton shower
is crucial especially to describe the jet-related observables. This is
done in two independent
studies~\cite{Frederix:2020jzp,Cordero:2021iau}, where the $t \bar t
W$ production is evaluated at NLO including the $\mathcal O
(\alpha_S^3\alpha)$ and $\mathcal O (\alpha_S\alpha^3)$ corrections,
followed by LO decays of the $t \bar t W$ resonances and matched to
the parton shower. The $\mathcal O (\alpha_S\alpha^3)$ contributions
do not include EW corrections from $\mathcal O (\alpha_S\alpha^2)$,
since the latter is zero due to the colour structure. This allows one
to treat them as pure QCD corrections to $\mathcal O (\alpha^3)$ and
therefore match them to the parton shower within the standard
frameworks~\cite{Frixione:2002ik,Nason:2004rx,Alioli:2010xd,Alwall:2014hca,Sherpa:2019gpd}. Apart
from their effects to differential distributions,
Ref.~\cite{Frederix:2020jzp} discusses their spin structure and
Ref.~\cite{Cordero:2021iau} shows their dependence on the matching to
the shower parameters.

Despite the continuous improvements on the $t \bar t W$ calculation,
the remaining tensions with respect to the experimental data show that
this process is theoretically not under complete control. A study on
the structure of the higher order contributions in $t \bar t W$ argues
that an extra $\sim\!\!10\%$ increase of the cross section is expected
from $\mathcal O (\alpha_S^4\alpha)$
corrections~\cite{vonBuddenbrock:2020ter}. In the absence of an NNLO
$t \bar t W$ calculation the multi-jet merging at NLO will capture
parts of these contributions, since the contributions that include
hard non-logarithmically enhanced radiation can be included at NLO
accuracy. Pragmatically, at NLO QCD the real-emission radiation can be
classified either as a QCD-jet ($j_{_{\rm QCD}}$) or as a Weak-jet
($j_{_{\rm Weak}}$). With the former we mean a jet that is attached to
a QCD vertex, while a Weak-jet is attached via an EW vertex to the $W$
boson, in a suitable (quasi)-collinear and/or soft limit. The parton
shower includes the $j_{_{\rm QCD}}$ emissions via QCD splitting
functions to $pp \rightarrow t \bar t W$, but not the $j_{_{\rm
    Weak}}$ ones. As a result, at a LO merging, the $t \bar t W
j_{_{\rm Weak}}$ contributions are omitted below the chosen merging
scale. Similarly, at a NLO merging, these contributions are evaluated
only at LO below the merging scale and at NLO above it. The presence
of the Weak-jet configurations in the $t \bar t W j$ contributions in
combination with the opening of the $qg$-initiated diagrams at NLO,
creates discontinuities in the characteristic differential jet-resolution 
($\sqrt{y_{ij}}$) distributions and the transverse momenta of the jets. The same
feature extends also for the $t \bar t W j j$ events. Within
{\sc\small MadGraph5\_aMC@NLO}~\cite{Alwall:2014hca,Frederix:2018nkq}
at LO this issue is resolved by excluding these configurations from
the merging and treat them as independent finite contributions. At NLO
the separation of these configurations followed by the NLO matching
and the different NLO matrix-element merging is not done up to now.

The merging procedure within {\sc\small MadGraph5\_aMC@NLO} is
upgraded to NLO in QCD with the {\sc\small FxFx}
framework~\cite{Frederix:2012ps}. In this project we extend the
{\sc\small FxFx} NLO QCD merging framework in order to correctly take
into account the $t \bar t W j_{_{\rm Weak}}$ contributions. We then
perform the NLO QCD merging for $t \bar t W$ up to two jets
and the matching to the parton shower
within the {\sc\small PYTHIA8} framework~\cite{Sjostrand:2014zea} via
the MC@NLO matching
method~\cite{Frixione:2002ik,Frixione:2003ei}. Following what is done
in Ref.~\cite{Frederix:2020jzp}, we include the $\mathcal O
(\alpha_S\alpha^3)$ subleading EW corrections. We calculate the
inclusive cross section including all these contributions and present
their effects on $t \bar t W$ differential distributions. We further
proceed to LO decays within the narrow-width-approximation (NWA),
maintaining the tree-level spin correlations focusing on the emerging
multi-lepton signatures. We study these effects at the cross-section
and differential-distribution level in the fiducial region.

The structure of this paper is the following. In Sec.~\ref{sec:theor_frame} we discuss the problems emerging to the merging procedure with the presence of the Weak-jets and describe the solution we implement within the {\sc\small FxFx} framework. We show how our implementation gives the correct differential jet-resolution and jet transverse-momentum distributions. In Sec.~\ref{sec:input_par} we show the input parameters and setup of our calculation. In Sec.~\ref{sec:results} we present our results at the cross section and differential distributions keeping the $t\bar t W$ stable and at the multi-lepton signatures after the $t\bar t W$ decay. In Sec.~\ref{sec:concl} we discuss our conclusions and outlook.

\section{Theoretical framework}
\label{sec:theor_frame}
In the real emission diagrams entering the $\mathcal O
(\alpha_S^3\alpha)$ corrections, one can distinguish two types of
contributions: the contributions where the extra emission is attached
to a QCD vertex and the ones where it is attached to an EW
vertex. While for general kinematic configurations this distinction is
somewhat ambiguous, in a suitable soft and/or (quasi)-collinear limit
it is well-defined\footnote{As discussed below, away from the strict
limits we use a clustering algorithm to determine the
classification.}.
\begin{figure}[h!]
\centering
\includegraphics[width=0.425\textwidth]{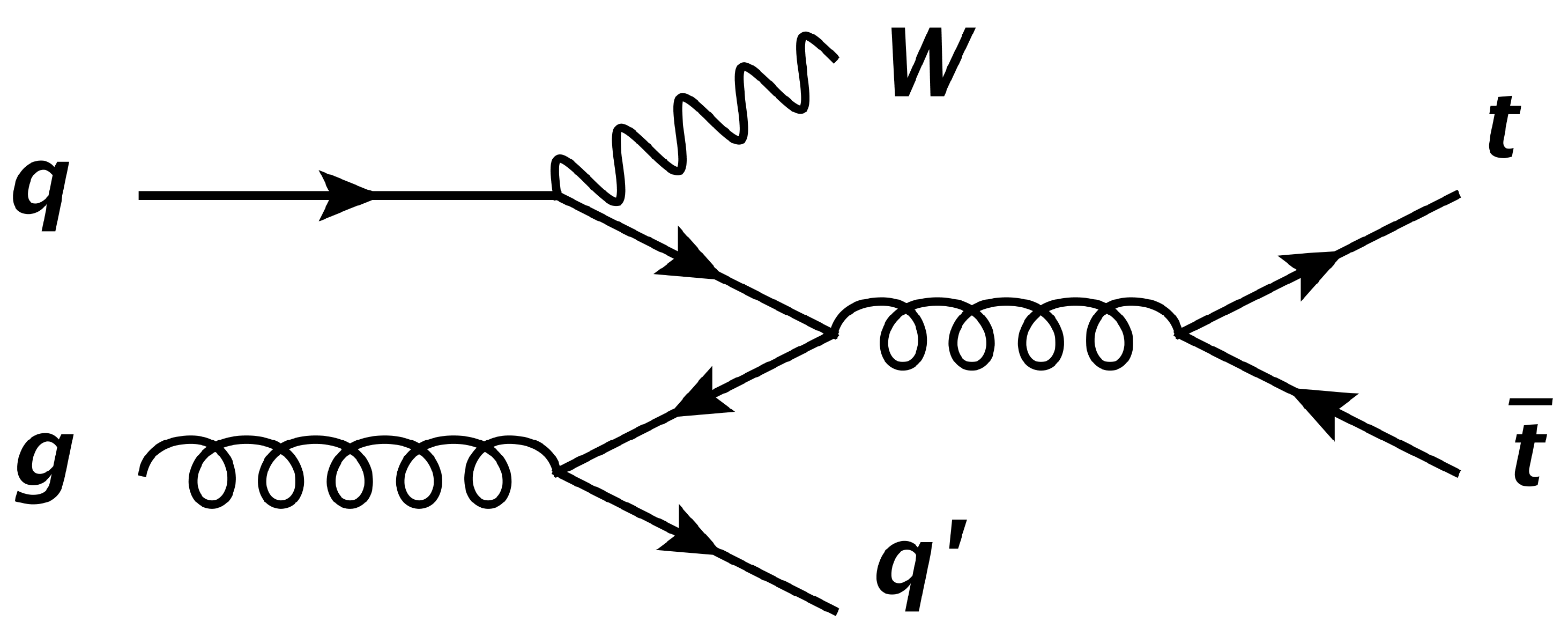} $\;\;\;\;\;\;\;\;$
\includegraphics[width=0.425\textwidth]{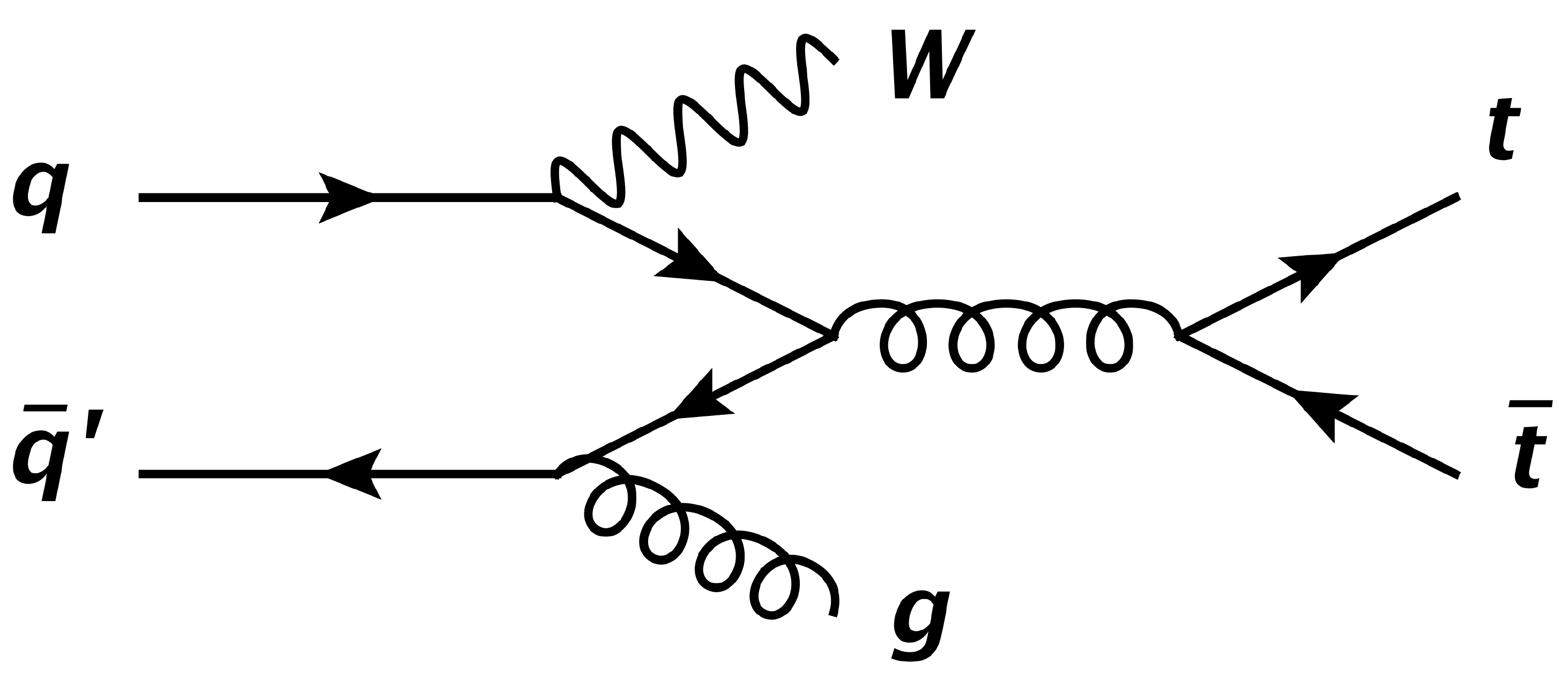} \\[1em]
\includegraphics[width=0.405\textwidth]{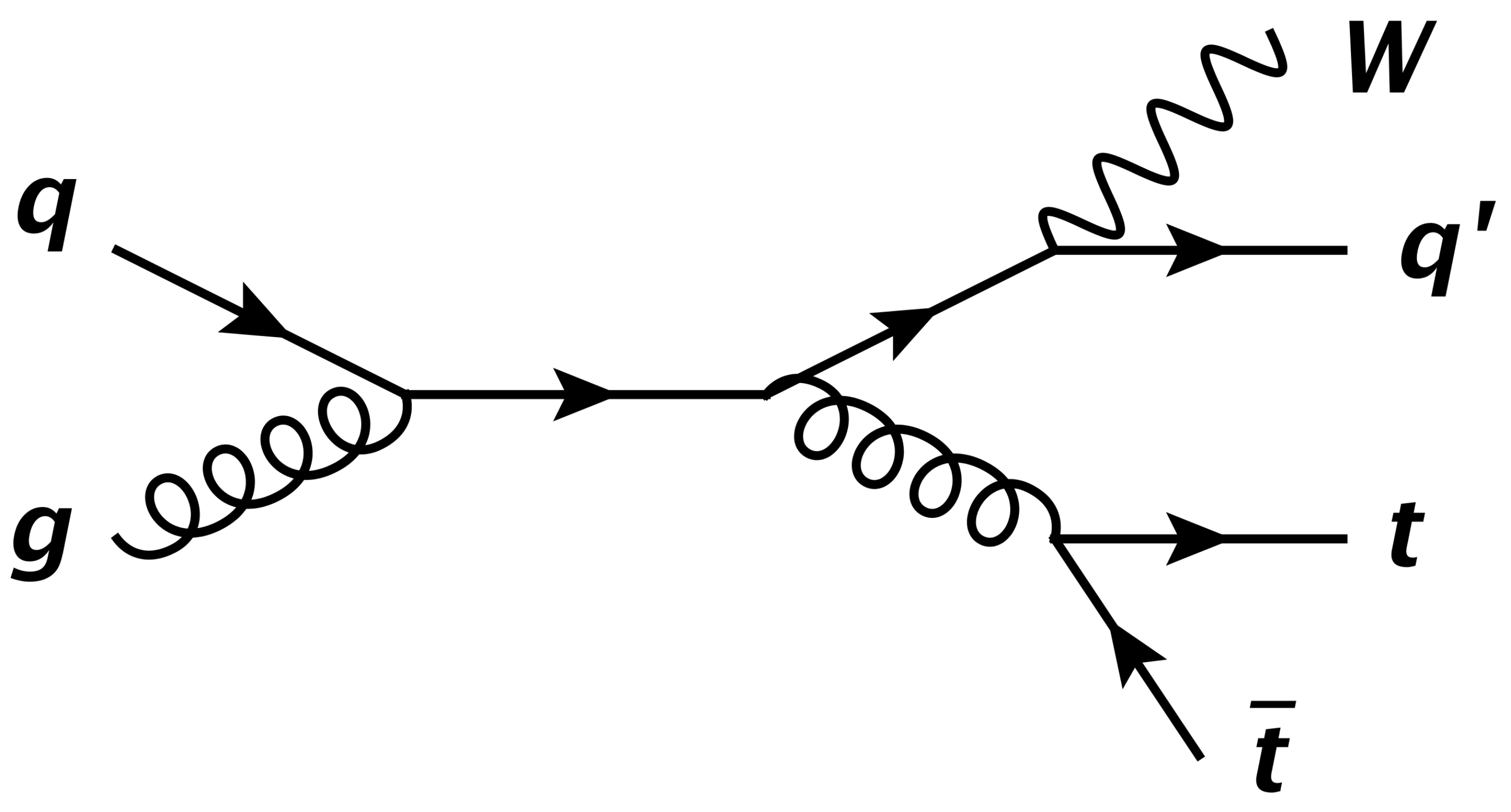} $\;\;\;\;\;\;\;\;$
\includegraphics[width=0.405\textwidth]{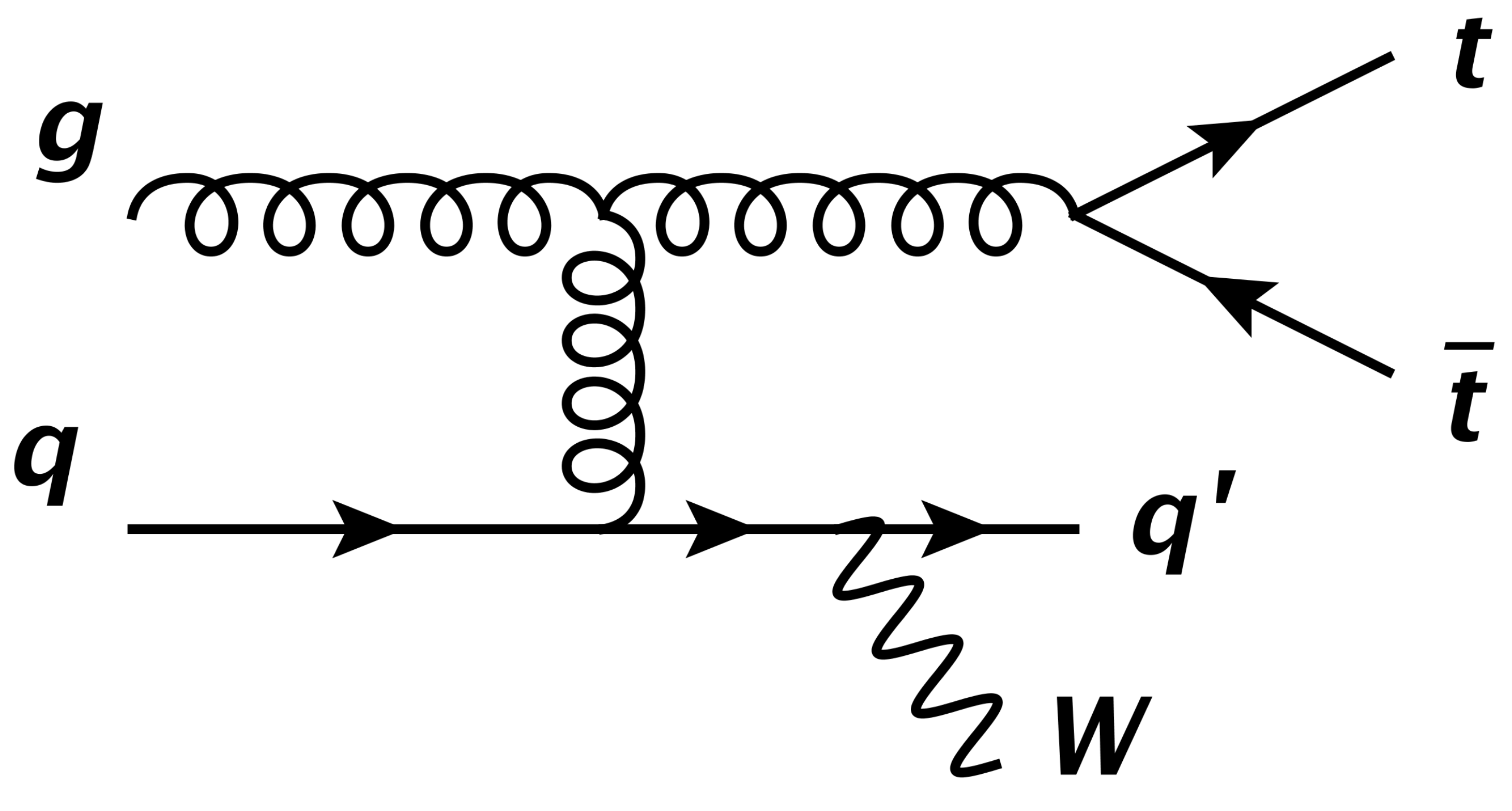}
\caption{Representative Feynman diagrams entering the $\mathcal O (\alpha_S^3\alpha)$ perturbative order. Different configurations, where the extra emission is a $j_{_{\rm QCD}}$ (upper diagrams) and a $j_{_{\rm Weak}}$ (lower diagrams).}
\label{fig:ttxWj_feyn}
\end{figure}
\noindent
For the rest of this paper we denote the former as $t \bar t W
j_{_{\rm QCD}}$ and the latter as $t \bar t W j_{_{\rm Weak}}$, with
$j_{_{\rm QCD}}$ and $j_{_{\rm Weak}}$ being the corresponding extra
emissions. In Fig.~\ref{fig:ttxWj_feyn} we show representative
diagrams of these two configurations. One should note that the two
configurations enter the same perturbative order, therefore the $t
\bar t W j_{_{\rm Weak}}$ contributions are suppressed neither
perturbatively nor by any kinematic reason. Furthermore for the
$\mathcal O (\alpha_S^3\alpha)$ perturbative order, the $j_{_{\rm
    Weak}}$ emissions appear only in the $qg$-induced contributions,
whereas the $j_{_{\rm QCD}}$ ones appear in both the $q\bar q$- and
the $qg$-induced contributions.

In a $t \bar t W$ {\sc\small FxFx} merging up to one jet, the $0$-jet
event sample will include the LO $pp \rightarrow t \bar t W j$
contributions, as part of the real emission corrections to $pp
\rightarrow t \bar t W$. This event sample will contribute to the
phase-space region below the merging scale. The $1-$jet event sample
will include these contributions at NLO in QCD, to which the $pp
\rightarrow t \bar t W j j$ contributions open up for the first time
as real emission corrections. This sample will contribute the
phase-space region above the merging scale. The separation of the
phase space region can be seen in the 1 to 0 jet resolution distribution, $\sqrt{y_{01}}$, and the
leading jet transverse-momentum differential distributions. We first
employ a $t \bar t W$ {\sc\small FxFx} merging up to one jet, without
our proposed implementation, in order to point out the emerging
problem. Choosing a merging scale of $\mu_Q=150$ GeV and keeping the
$t \bar t W$ resonances stable, we merge within the {\sc\small FxFx}
framework the $t \bar t W$@NLO + $t \bar t Wj$@NLO matrix elements and
match them to the parton shower wihtin the {\sc\small PYTHIA8}
framework.
\begin{figure}[b!]
\centering
\includegraphics[width=1\textwidth]{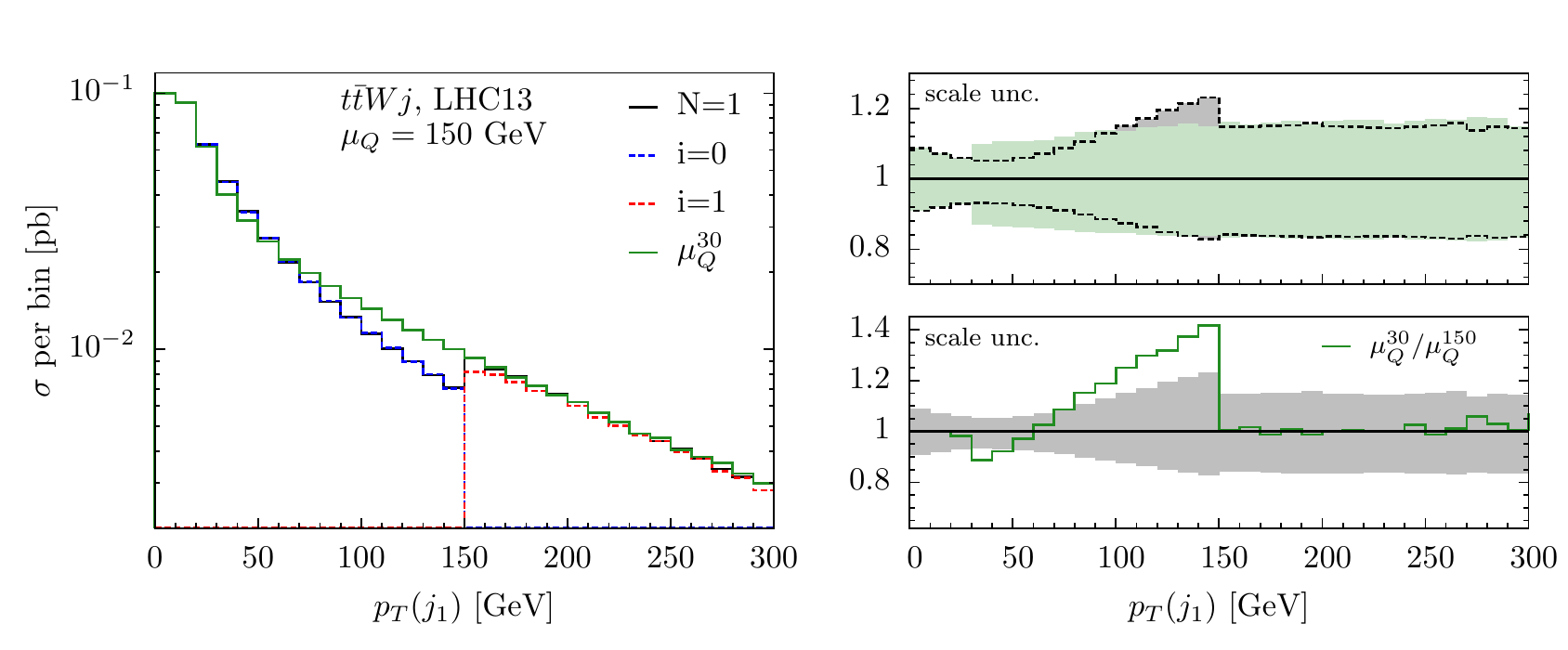}
\caption{Transverse-momentum differential distribution of the leading jet in a $t \bar t W$ {\sc\small FxFx} merged sample using $\mu_Q=150$ GeV as merging scale (left plot). Comparison with $\mu_Q=30$ GeV merging scale regarding the central value and the scale uncertainties (right ratio plots).}
\label{fig:def_Qcut}
\end{figure}
\noindent
In the left plot of Fig.~\ref{fig:def_Qcut} we show the
transverse-momentum differential distribution of the leading jet. We
can first see the clear phase space separation between the $0-$jet
($i=0$, dashed blue line) and the $1$-jet ($i=1$, dashed red line) at
150 GeV. Furthermore, the merged sample ($N=1$, black solid line)
shows a significant discontinuity at the merging point. One can easily
deduce that varying the merging scale in the range e.g.~$100-200$ GeV
will displace accordingly the discontinuity and therefore introduce a
spurious dependence on the integrated cross section. This dependence
is shown numerically in detail and discussed in
Ref.~\cite{vonBuddenbrock:2020ter}. The reason for this discontinuity
is the two different types of real-emission contributions depicted in
Fig.~\ref{fig:ttxWj_feyn}. Below the merging scale, all the
$t\bar{t}Wj$ contributions are evaluated at LO within the matrix
element. In the $0-$jet sample, after matching to the parton shower,
the Sudakov suppression is imposed by vetoing emissions from the hard
scale ($Q_h$) of the event down to the merging scale. The Sudakov
factors include the real unresolved and soft virtual corrections
resummed at LL and furthermore the parton shower will add emissions to
the $0-$jet sample below the merging scale. The effects introduced
this way mimic the higher order corrections for the $t \bar t W
j_{_{\rm QCD}}$ contributions but not for the $t \bar t W j_{_{\rm
    Weak}}$ ones. The reason for this difference is that the $t \bar t
W j_{_{\rm Weak}}$ emissions are not reproduced by the parton shower.

The apparent solution is to set a very low merging scale, e.g.~$\mu_Q=30$ GeV (green solid line). This solution hides the discontinuity in a phase-space region dominated by the parton-shower effects and describes the vast majority of the phase space with the $1-$jet multiplicity sample. However this introduces large logarithms $L \sim \log(Q_h/\mu_Q)$ and reduces the accuracy of the distribution, as discussed in detail in Ref.~\cite{Hamilton:2012rf}. In the right ratio plots of Fig.~\ref{fig:def_Qcut} we show the comparison between the two merging-scale choices of 150 and 30 GeV. In the lower inset we can see the ratio of the central values and the displacement of the discontinuity from 150 to 30 GeV. In the upper inset we show that the discontinuity is also present in the range of the scale variation before and after the selected merging scale in both choices.

In our implementation within the {\sc\small FxFx} framework we follow
the merging procedure that is presented in
Ref.~\cite{Frederix:2012ps}. This procedure builds upon the {\sc\small
  MiNLO} approach~\cite{Hamilton:2012np}, which is also described in
detail in Sec.~1.4 of Ref.~\cite{Moretti:2016jnv}. Respecting the
flavour and colour structure we cluster with a $k_T$ algorithm, but we
do not restrict ourselves to QCD partons. This way we can separate the
QCD from the EW vertices by considering the clustering tree and
therefore distinguish the $t \bar t W j_{_{\rm Weak}}$ from the $t
\bar t W j_{_{\rm QCD}}$ contributions. Contributions that are flagged
as including a $j_{_{\rm Weak}}$ parton are not divergent in the limit
$p_T(j_{_{\rm Weak}})\to 0$ since they are regulated by the massive
vector boson. Therefore they can be described by the matrix element in
the full phase space. Hence, the $j_{_{\rm Weak}}$ partons are labeled
as such in the event file and eventually excluded from the matching
procedure within {\sc\small PYTHIA8}. On the other hand, the treatment
of the $t \bar t W j_{_{\rm QCD}}$ contributions and the $j_{_{\rm
    QCD}}$ emitted partons are not altered with respect to the default
{\sc\small FxFx} framework. The effects of this implementation in the
differential distributions, like the one in Fig.~\ref{fig:def_Qcut}
will be shown and discussed in detail in Sec.~\ref{sec:val}, after we
introduce the input parameters and the calculation setup.

\section{Calculation setup - input parameters}
\label{sec:input_par}

In this section we will focus on more technical aspects of the
calculation and the input parameters. We will start with the central
renormalisation and factorisation scale definitions. In the merged
sample we adopt the renormalisation scale definition of
Ref.~\cite{Frederix:2012ps}, which is in agreement with the definition
within the {\sc\small MiNLO} framework. The $t \bar t W$ Born
contribution to the $0-$jet event sample is of the order of $\mathcal
O (\alpha_S^2\alpha)$ and any configuration, regardless the
perturbative order and the jet multiplicity it belongs, can have a
number of $n$ light-parton QCD clusterings associated to $d_n$
distance measure clustering scales. For the real emission
contributions to each sample we omit from this counting the clustering
with the lowest $d$ scale in order to maintain the inclusive
integration over the extra emission. The renormalisation scale is then
defined as

\begin{equation}
\mu_r^0 = \left[ \left(\mu_Q^h \right)^2 \prod_{i = 1}^{n} d_{i} \right]^{1/(2+n)} \; ,
\label{eq:ren_scale}
\end{equation}
\noindent
where $\mu_Q^h$ is the scale associated with the hard $Q_h$ scale of the process. In our default calculation setup $\mu_Q^h$ is evaluated in an event by event basis as $\mu_Q^h = {\rm max}(Q_h,d_1)$, with $d_1$ being the hardest QCD clustering scale. For our investigation on the scale dependence of our results we further employ different $\mu_Q^h$ functional forms. In all cases, for the merged sample the central factorisation scale is set in an event by event basis as 
\begin{equation}
\mu_f^0 = \left \{ \begin{aligned} & d_n \text{ for } n \neq 0 \\[1ex] & \mu_r^0 \text{ for } n=0 \end{aligned}  \right. \; ,
\label{eq:fac_scale}
\end{equation}
\noindent
where $d_n$ is the lowest among the $d_i$ clustering scales. Regarding all our non-merged stand-alone predictions it is $\mu_f^0 = \mu_r^0$. For the scale dependence investigation we will use the four different renormalisation scale functional forms shown in Tab.~\ref{tab:scales}.
\begin{table}[h]
\small
\begin{center}
\renewcommand{\arraystretch}{1.5}
\begin{tabular}{c c | c c }
\hline
\multicolumn{2}{c}{non-merged} & \multicolumn{2}{c}{merged} \\
\hline
  &  &  &  \\
$\mu_r^0 = \left \{ \begin{aligned} & \dfrac{H_T'}{2}= \dfrac{\sum_i m_{T,i}}{2}, i=t,\bar t, W, (j) \; {(\rm default)} \\[1ex] & \dfrac{H_T}{2}= \dfrac{\sum_i m_{T,i}}{2}, i=t,\bar t, W \\[1ex] & \dfrac{Q}{2}= \dfrac{M_{\rm inv}(t\bar t W)}{2} \\[1ex] & \dfrac{M}{2}= \dfrac{2M_t+M_W}{2} \end{aligned}  \right.$  &  & $\mu_Q^h = \left \{ \begin{aligned} & {\rm max}(Q_h,d_1) \; {\rm (default)} \\[1ex] & \dfrac{H_T}{2}= \dfrac{\sum_i m_{T,i}}{2}, i=t,\bar t, W \\[1ex] & \dfrac{Q}{2}= \dfrac{M_{\rm inv}(t\bar t W)}{2} \\[1ex] & \dfrac{M}{2}= \dfrac{2M_t+M_W}{2} \end{aligned}  \right.$ &  \\
  &  &  &  \\
\hline
\end{tabular}
\end{center}
\caption{Different functional forms for the central value of the renormalisation scale used for the scale dependence study.}  
\label{tab:scales}
\end{table}
\noindent
For a given $\mu_Q^h$, one can derive via Eq. \ref{eq:ren_scale} the corresponding $\mu_r$ for the merged samples. The first line in Tab.~\ref{tab:scales} corresponds to the default scales in {\small \sc MadGraph5\_aMC@NLO}. Apart from the default values we choose the specific functional forms, since they are the ones used in the two NLO+NNLL studies on $t \bar t W$. The scales $H_T/2$ and $Q/2$ are the ones that are used and eventually combined in Ref.~\cite{Broggio:2019ewu}. In the study of Ref.~\cite{Kulesza:2020nfh} the fixed scale $M/2$ is also included on top of the other two dynamical choices. In all cases we derive the scale uncertainties via the usual $9-$point variation of the renormalisation and factorisation scales in the range $\{\frac{1}{2}\, \mu,2\mu\}$. Regarding the rest of our input, we use the 5 Flavour-Scheme (FS) with the corresponding NLO NNPDF31 PDF sets~\cite{NNPDF:2017mvq} and the following parameters
\begin{equation}
  \begin{tabular}{l l l}
    $M_t = 173.34 \; {\rm GeV}$, &  $M_Z = 91.1876\; {\rm GeV}$, &  $M_\tau = 1.777\; {\rm GeV}$,  \\
    $\Gamma_t = 1.49150 \; {\rm GeV}$, & $\Gamma_Z = 2.4414\; {\rm GeV}$, & $\Gamma_W = 2.0476\; {\rm GeV}$, \\
    $\alpha_{EW} = 1/132.232$, & $G_\mu = 1.16639 \times 10^{-5}\; {\rm GeV}^{-2}$, & $(\Rightarrow M_W = 80.385\; {\rm GeV})$.
  \end{tabular}
  \label{eq:input_par}
\end{equation}
\noindent

Our results consist of two main parts. The first part is the production level, where the $t$, $\bar t$ and $W$ resonances are kept stable. We calculate the $pp \rightarrow t \bar t W$ process at LO QCD, NLO QCD and after NLO merging up to one (FxFx@1J) and  two (FxFx@2J) jets. The matrix elements are matched to the parton shower via {\sc\small PYTHIA8} and we will show results at the cross section and differential level. For this part we do not include hadronisation after the parton shower. For the jet identification we use a $k_T$ jet algorithm~\cite{Catani:1993hr} with $R=1$. For the cross sections we further add the subleading EW corrections of $\mathcal O (\alpha^3)$ and $\mathcal O (\alpha_S\alpha^3)$ ($\textrm{NLO}_\textrm{EW}^\textrm{sub}$). For more details regarding their inclusion and matching to the parton shower we refer the reader to Refs.~\cite{Frederix:2020jzp,Cordero:2021iau}. Concerning the leading EW corrections of $\mathcal O (\alpha_S^2\alpha^2)$ ($\textrm{NLO}_\textrm{EW}^\textrm{lead}$), the matching to the parton shower is not yet possible, and we include these contributions only in our inclusive cross-section predictions, by calculating them at fixed order. Regarding the EW corrections we omit the $\mathcal O (\alpha^4)$ corrections, since they are at the permille level w.r.t.~the LO QCD~\cite{Frederix:2017wme}. The various perturbative orders entering our calculation are
\begin{align}
  \textrm{LO}_\textrm{QCD} &= t \bar t W @ \mathcal O (\alpha_S^2\alpha) \; , \nonumber \\
  \textrm{NLO}_\textrm{QCD}& = t \bar t W @ \mathcal O (\alpha_S^2\alpha, \alpha_S^3\alpha) \;,  \nonumber \\
  \textrm{NLO}_\textrm{EW}^\textrm{lead} &= t \bar t W @ \mathcal O (\alpha_S^2\alpha^2) \; , \nonumber \\
  \textrm{NLO}_\textrm{EW}^\textrm{sub} &= t \bar t W @ \mathcal O (\alpha^3, \alpha_S\alpha^3) \;,  \nonumber \\
\textrm{FxFx@1J} &= t \bar t W @ \mathcal O (\alpha_S^2\alpha, \alpha_S^3\alpha) + t \bar t W j @ \mathcal O (\alpha_S^3\alpha, \alpha_S^4\alpha) \; , \nonumber \\
\textrm{FxFx@2J} &= t \bar t W @ \mathcal O (\alpha_S^2\alpha, \alpha_S^3\alpha) + t \bar t W j @ \mathcal O (\alpha_S^3\alpha, \alpha_S^4\alpha) + t \bar t W j j @ \mathcal O (\alpha_S^4\alpha, \alpha_S^5\alpha) \; .
\label{eq:orders}
\end{align}
\noindent
We note that despite the fact that the $qg$ and $gg$ contributions open up for the first time at $\mathcal O (\alpha_S^3\alpha)$ and $\mathcal O (\alpha_S^4\alpha)$, they are not IR finite (except the ones that include the $j_{_{\rm Weak}}$'s, as we have discussed in Sec.~\ref{sec:theor_frame}). The main missing ingredient in order to obtain fully the $t \bar t W$ at NNLO QCD precision (up to $\mathcal O (\alpha_S^4\alpha)$) is the 2-loop virtual $q\bar q$ contributions. The soft-gluon resummation performed in Refs.~\cite{Broggio:2019ewu,Kulesza:2020nfh} shows that the resummed up to all orders soft virtual and real emission corrections to $q \bar q-$induced diagrams have a few percent effect on the complete NLO prediction for the scales shown in Tab.~\ref{tab:scales}. In view of the pure QCD nature of the resummation for $t \bar t W$, the NLO merging is the only consistent framework to date, where both the $qg$ and $gg$ contributions can be included in the calculation taking into account the emerging Weak-jet configurations. For these reasons the merging procedure for $t \bar t W$, on top of its usual purpose on improving any jet-related observable, becomes interesting also at the inclusive level. We use this first part of the calculation in order to study the scale dependence on various levels of accuracy (Sec.~\ref{sec:scale}). We then provide our cross section predictions (Sec.~\ref{sec:xsec}) and we show differential distributions for the $t$, $\bar t$ and $W$ particles (Sec.~\ref{sec:diff_distrib}).

The second part of our results are at the decay level, where we decay the $t$, $\bar t$ and $W$ resonances in the narrow-width approximation using {\small \sc MadSpin}~\cite{Artoisenet:2012st}, keeping the tree-level spin correlations. We allow for all possible decays in order to be able to provide predictions for various multi-lepton signatures. For this part the parton shower is followed by the hadronisation in order to be more realistic to the final signatures. For the selections and cuts we follow the settings of the analysis in Ref.~\cite{ATLAS:2019nvo}, which are the ones presented in detail and utilised in Ref.~\cite{Frederix:2020jzp}. The decays of the $\tau$ leptons take place within {\small \sc PYTHIA8} and our final jet and lepton definitions and signal-region selections are\footnote{for more details on the signal region selections and the particle identification we refer the reader to Ref.~\cite{Frederix:2020jzp}.}:
\begin{align}
{\rm Jets:}\,\, &\; \textrm{anti-$k_T$~\cite{Cacciari:2008gp}} \; , \; R=0.4 \; , \; |\eta(j)| \le 2.5 \; , \; p_T(j) \ge 25 \; {\rm GeV} \nonumber \\
{\rm Leptons:}\,\, &\; |\eta(\ell)| \le 2.5 \; , \; p_T(\ell) \ge 10 \; {\rm GeV} \nonumber \\
{\rm Signal \; regions}\,\,& (n_j\ge 2, n_{b_j} \ge 1): \; {\rm same \; sign \; dilepton} \rightarrow 2ss\ell \; , \; {\rm trilepton} \rightarrow 3\ell \;.
\label{eq:id}
\end{align}
\noindent
Our calculation does not include efficiencies on particle identification or misidentifications between light-jets, $b-$jets and leptons. The $b-$jets are defined as jets containing at least one $B$ hadron. For this second part we focus on the cross section of the various multi-lepton signatures at the fiducial region and jet-related differential distributions in order to point out the effects of the {\small \sc FxFx} merging (Sec.~\ref{sec:multilep}).

\section{Results}
\label{sec:results}

In this section we will present our results. In Secs. \ref{sec:val} to \ref{sec:xsec} we use the production level setup, whereas in Sec.~\ref{sec:multilep} we move to the decay level. In each section we will specify which perturbative orders we include.

\subsection{Validation}
\label{sec:val}

\begin{figure}[t!]
\centering
\includegraphics[width=0.475\textwidth]{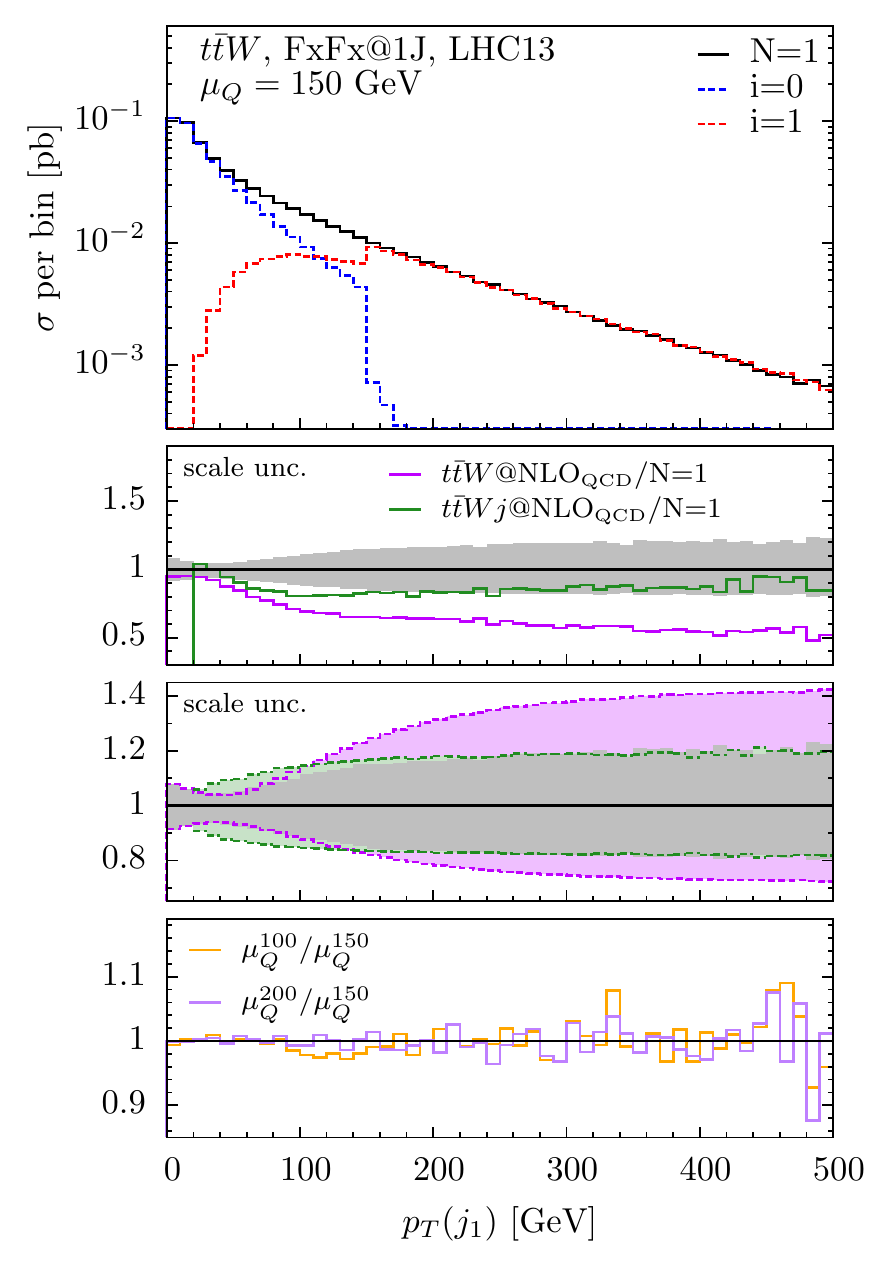}
\includegraphics[width=0.475\textwidth]{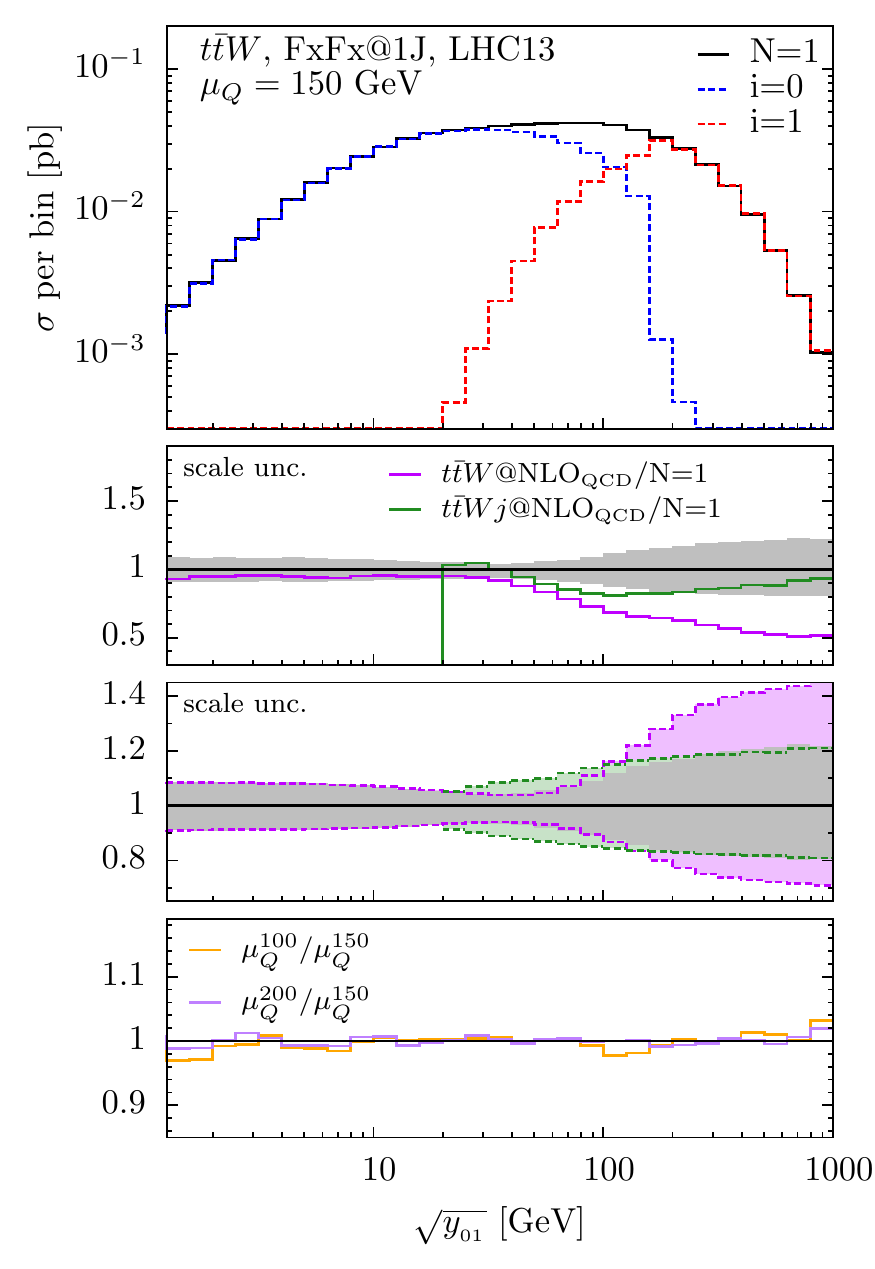} \\
\includegraphics[width=0.475\textwidth]{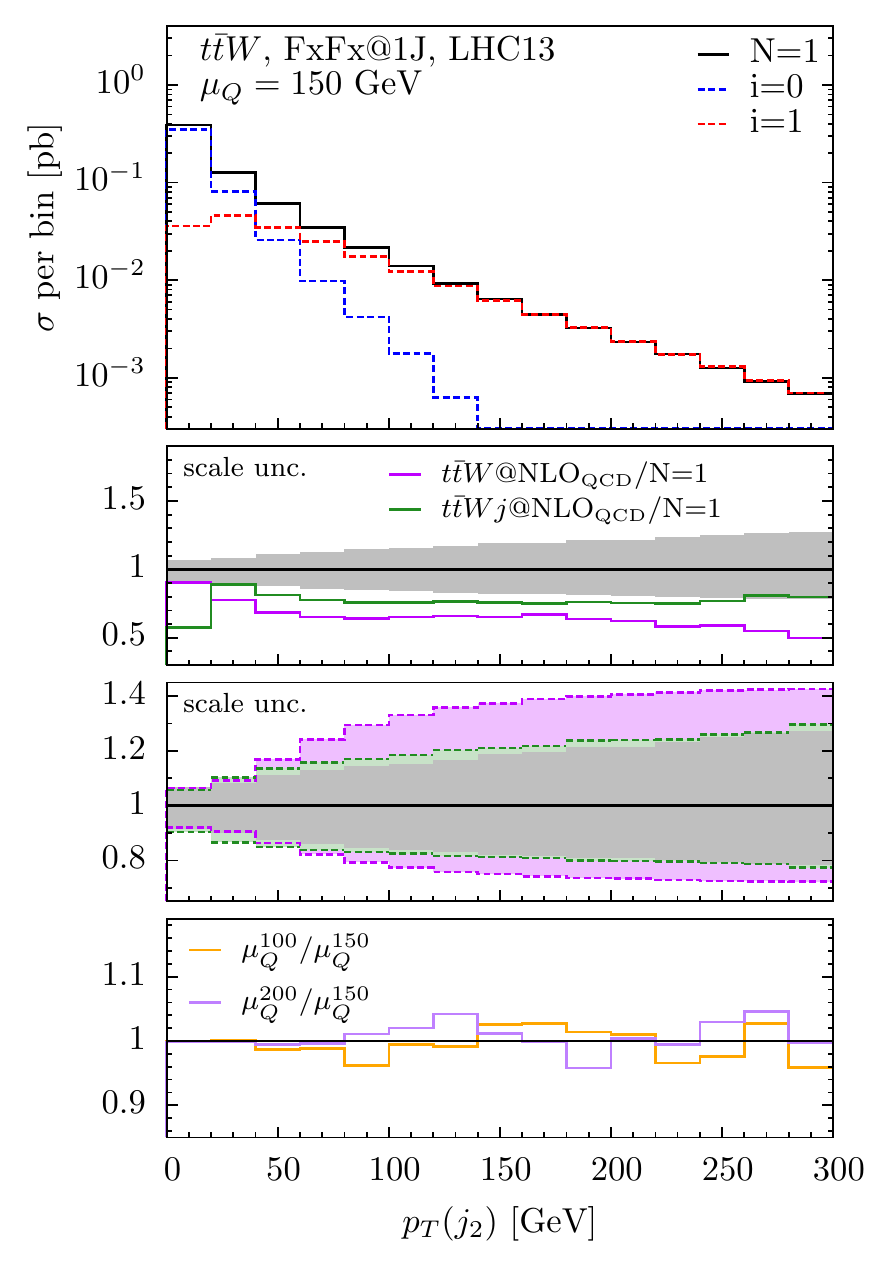}
\includegraphics[width=0.475\textwidth]{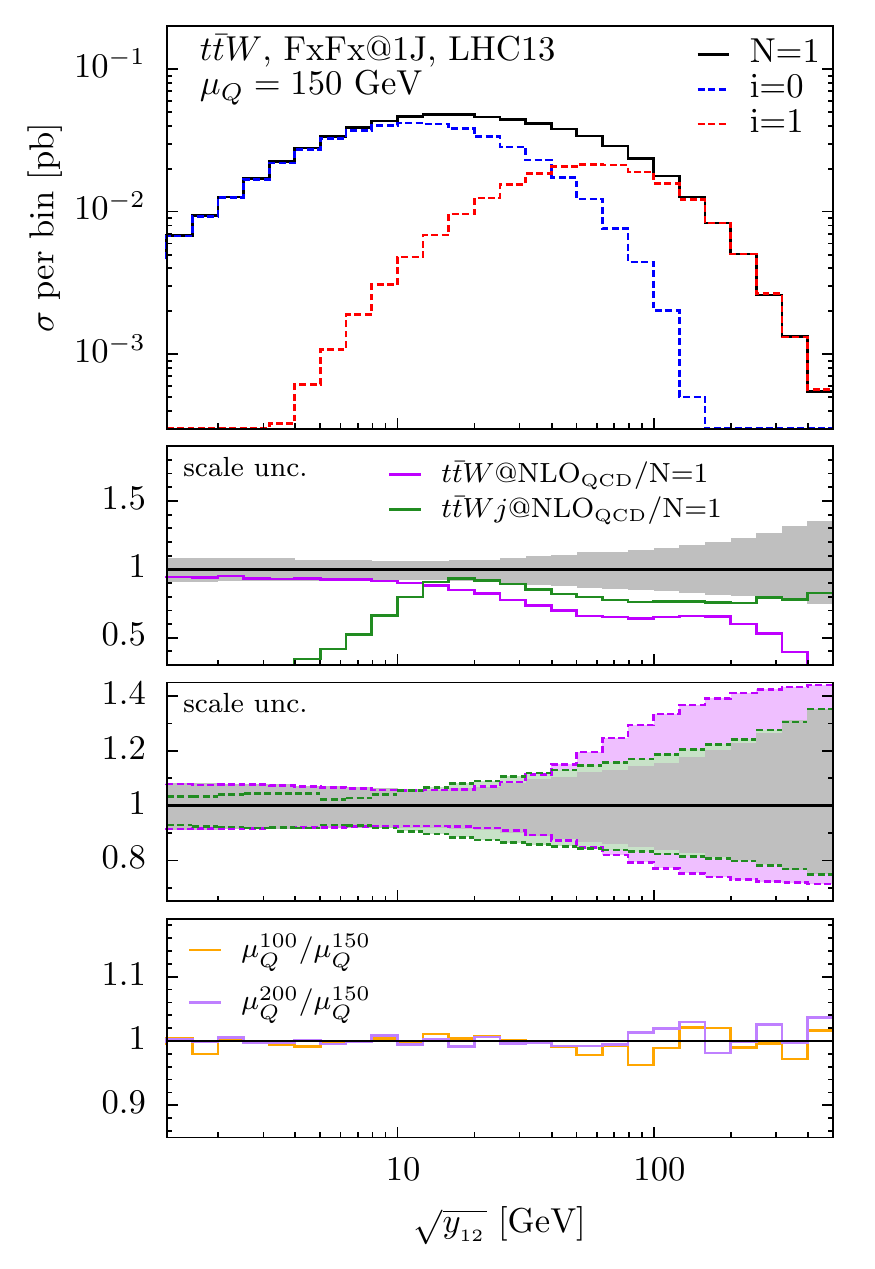}
\caption{Transverse-momentum differential distribution of the leading
  jet (left plot) and jet resolution $y_{01}$ distribution (right plot) for the
  $t \bar t W$ {\sc\small FxFx} merged sample using $\mu_Q=150$ GeV as
  merging scale.}
\label{fig:267_Qcut}
\end{figure}

In this section we intend to focus on the differential distributions
that are useful in order to validate the merging procedure. They are
the jet resolution and jet transverse-momentum distributions for our
{\small \sc FxFx} implementation. We further will demonstrate the
stability of our results under different choices for the merging scale
parameter. For this section we will use the default values for
renormalisation and factorisation scales, as defined in
Sec.~\ref{sec:input_par}. In Fig.~\ref{fig:267_Qcut} we show the $p_T$
distribution of the leading jet, $p_T(j_1)$, with the differential jet 
resolution from 1 to 0 jet, $\sqrt{y_{01}}$, (upper plots) and the $p_T$
distribution of the subleading jet with the $\sqrt{y_{12}}$ from 2 to 1 jets
(lower plots). In the main panel we show, similarly to
Fig.~\ref{fig:def_Qcut} the merged sample and separately the $0-$ and
$1-$ jet subsamples that take part in the merging. Comparing the
$p_T(j_1)$ distribution with the same plot in Fig.~\ref{fig:def_Qcut},
we can see the difference in the $1-$jet sample ($i=1$, dashed
red). This shows that with our implementation the $\mu_Q$ phase-space
limit is relaxed for the $t \bar t W j_{_{\rm Weak}}$ contributions
and they are allowed also below the merging scale. We can further see
that these contributions are finite, as expected, since the $j_{_{\rm
    Weak}}$ IR limit is regulated by the mass of the $W$. This
restores the continuum of the distribution that was lost in the plot
of Fig.~\ref{fig:def_Qcut}. The continuity in the phase-space limit is
apparent also in the $\sqrt{y_{01}}$ distribution in the upper right plot. In
the first inset we show the ratio of the stand-alone $t\bar t W$ and
$t\bar t Wj$ at NLO QCD over the merged sample (for the $t\bar{t}Wj$ stand-alone sample 
we require a generation cut of $p_T(j)>20$ GeV). In both these cases we
see that the low range regime of the merged sample agrees, within the
scale uncertainties, with the $t\bar t W$ at NLO QCD and the regime
after the merging scale phase-space limit agrees with the $t\bar t Wj$
at NLO QCD. The reason that the $t \bar t W@\textrm{NLO}_\textrm{QCD}$
prediction does not agree with the $\textrm{FxFx@1J}$ one up to the
merging-scale value is the fact that the $t \bar t W j_{_{\rm Weak}}$
contributions are evaluated at LO QCD in the former and at NLO QCD in
the latter throughout the full phase space. 
\noindent
In the second inset we compare the scale uncertainties of the $\textrm{FxFx@1J}$, $t \bar t W@\textrm{NLO}_\textrm{QCD}$ and $t \bar t Wj@\textrm{NLO}_\textrm{QCD}$ predictions. We can see that in the low energy regime ($p_T(j_1)\approx \sqrt{y_{01}} \lesssim 100$~GeV) the scale uncertainties of the $\textrm{FxFx@1J}$ prediction match the ones of the $t \bar t W@\textrm{NLO}_\textrm{QCD}$ one. After this value and for the rest of the phase space, the $\textrm{FxFx@1J}$ scale uncertainties match the ones of the $t \bar t Wj@\textrm{NLO}_\textrm{QCD}$ prediction. This inset reassures that there is no underestimation of the scale uncertainties of the merged sample throughout the phase space. In the lower plots there is an overlap between the $0-$jet and $1-$jet samples in the main panel. Similar behaviour is observed regarding the first two insets. In all cases, we further vary our choice for the merging scale by $\pm 50$~GeV and we show the ratio of these predictions over the $\mu_Q=150$ GeV value in the third inset.  We can see that the shapes of the distributions are not affected by this extended variation.

In order to further make certain that there is no dependence on the merging scale choice or underestimation of the scale uncertainties introduced by our choice of $\mu_Q=150$ GeV, we examine the total cross section prediction with various $\mu_Q$ choices. In Tab.~\ref{tab:mu_choice} we show the predictions in the range \{$25-150$ GeV\} with a step of $25$ GeV and the range \{$150-350$ GeV\} with a step of $50$ GeV. We can first appreciate that there is a stability in the scale variation regardless the merging-scale choice, even when choosing the very low value of 25 GeV. Furthermore in the whole range of \{$25-350$ GeV\} the variation of the merging scale corresponds to a cross-section deviation of less than $\sim \!\! 1.5\%$~\footnote{All these merging-scale choices of the implemented FxFx version do not yield similar results with the old version due to the different treatment of the $j_{_{\rm Weak}}$'s below and above the merging scale.}.

\begin{table}[h!]
\renewcommand{\arraystretch}{2.0}
\begin{center}
\resizebox{\columnwidth}{!}{
\begin{tabular}{c | c c c c c }
\hline
\multicolumn{6}{c}{$t \bar t W$, $\textrm{FxFx@1J}$} \\
\hline
$\mu_Q$ [GeV] & 25  & 50 & 75 & 100 & 125 \\
$\sigma$[fb] & $668.2(9)_{-77.5 (-11.6  \%)}^{+54.7 (+8.2  \%)}$ & $671.4 (8)_{-74.4 (-11.1  \%)}^{+60.0 (+8.9  \%)}$ & $673.6 (8)_{-71.9 (-10.7  \%)}^{+60.1 (+8.9  \%)}$ & $677.5 (6)_{-71.1 (-10.5  \%)}^{+60.8 (+9.0  \%)}$ & $677.2 (8)_{-69.5 (-10.3  \%)}^{+59.1 (+8.7  \%)}$ \\
\hline
$\mu_Q$ [GeV] & 150  & 200 & 250 & 300 & 350 \\
$\sigma$[fb] & $679.2(6)_{-69.7 (-10.3  \%)}^{+60.5 (+8.9  \%)}$ & $679.1(6)_{-69.5 (-10.2  \%)}^{+61.0 (+9.0  \%)}$ & $678.8(6)_{-69.6 (-10.3  \%)}^{+61.5 (+9.1  \%)}$ & $678.3(6)_{-69.6 (-10.3  \%)}^{+61.8 (+9.1  \%)}$ & $678.1(6)_{-69.7 (-10.3  \%)}^{+62.0 (+9.1  \%)}$ \\
\hline
\end{tabular}
}
\end{center}
\caption{Cross section comparisons for $t \bar t W$ using different merging scale choices. The error in parenthesis is the absolute MC-error on the last digit of the central value. The scale uncertainties are shown in the form of $ \pm \{\textrm{absolute}\}(\pm \{\textrm{relative in \%}\})$.}
\label{tab:mu_choice}  
\end{table}
\noindent

Having established the stability of our results regardless the choice of the merging scale, we use from now on the $\mu_Q=150$ GeV choice and we move to the next section with a study on differential distributions at the production level.

\subsection{Differential distributions}
\label{sec:diff_distrib}

\begin{figure}[t!]
\centering
\includegraphics[width=0.475\textwidth]{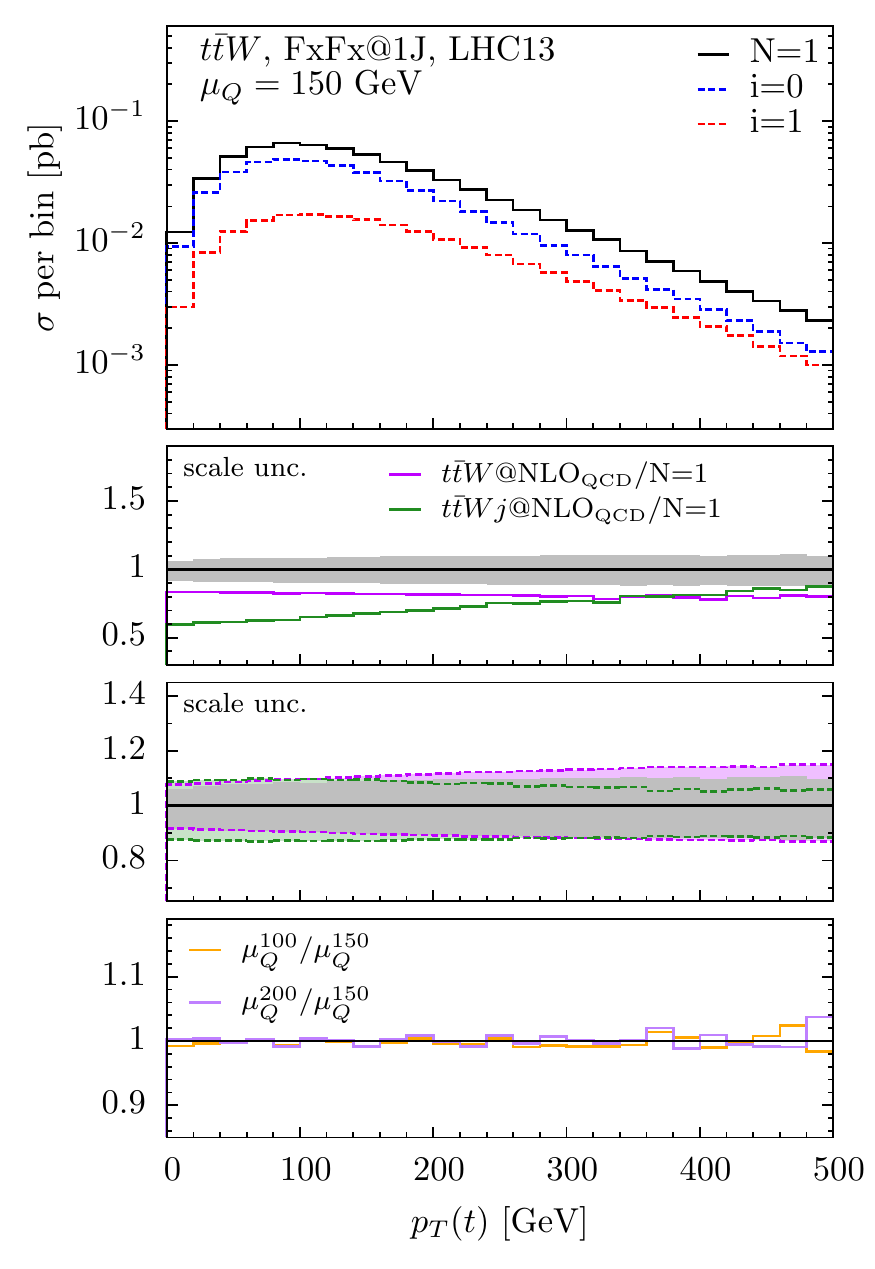}
\includegraphics[width=0.475\textwidth]{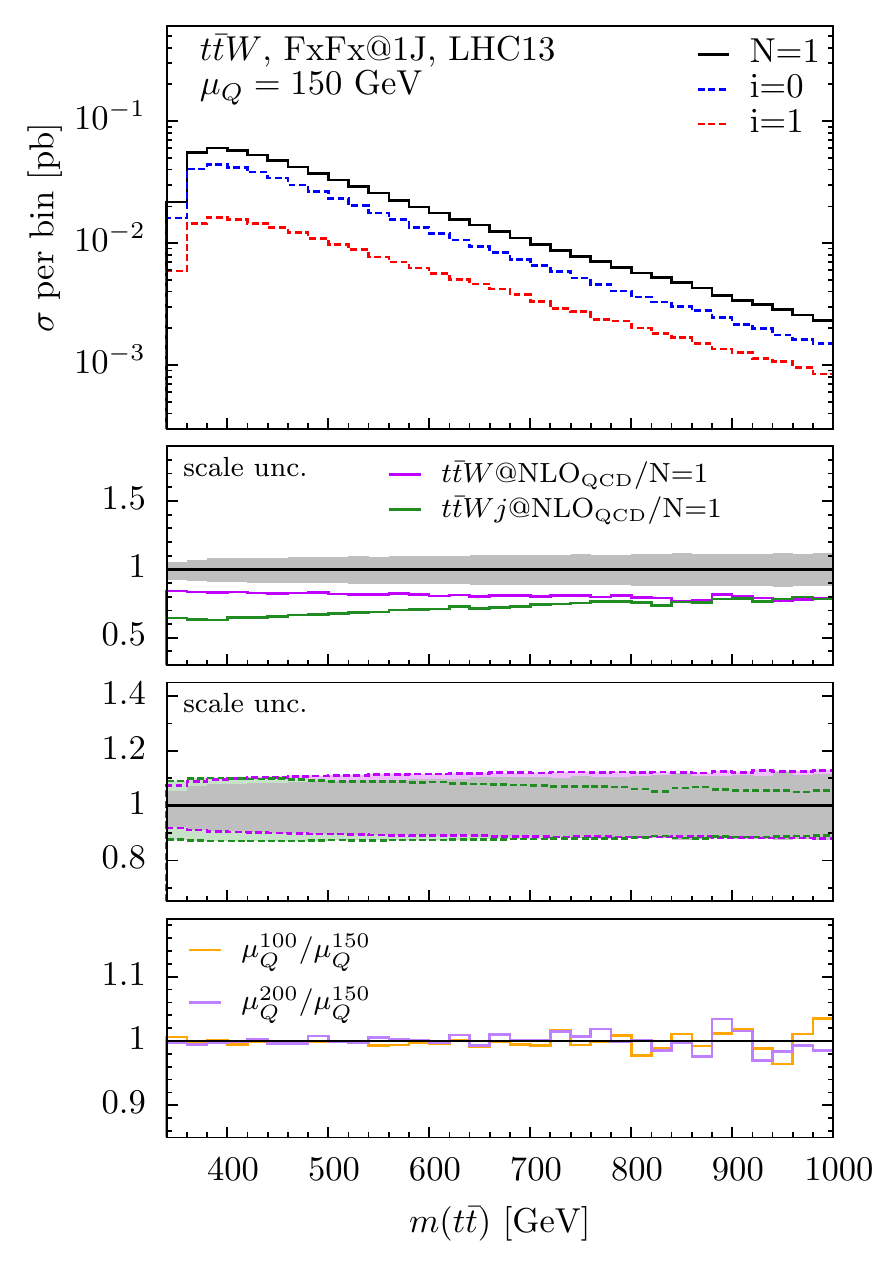} \\
\includegraphics[width=0.475\textwidth]{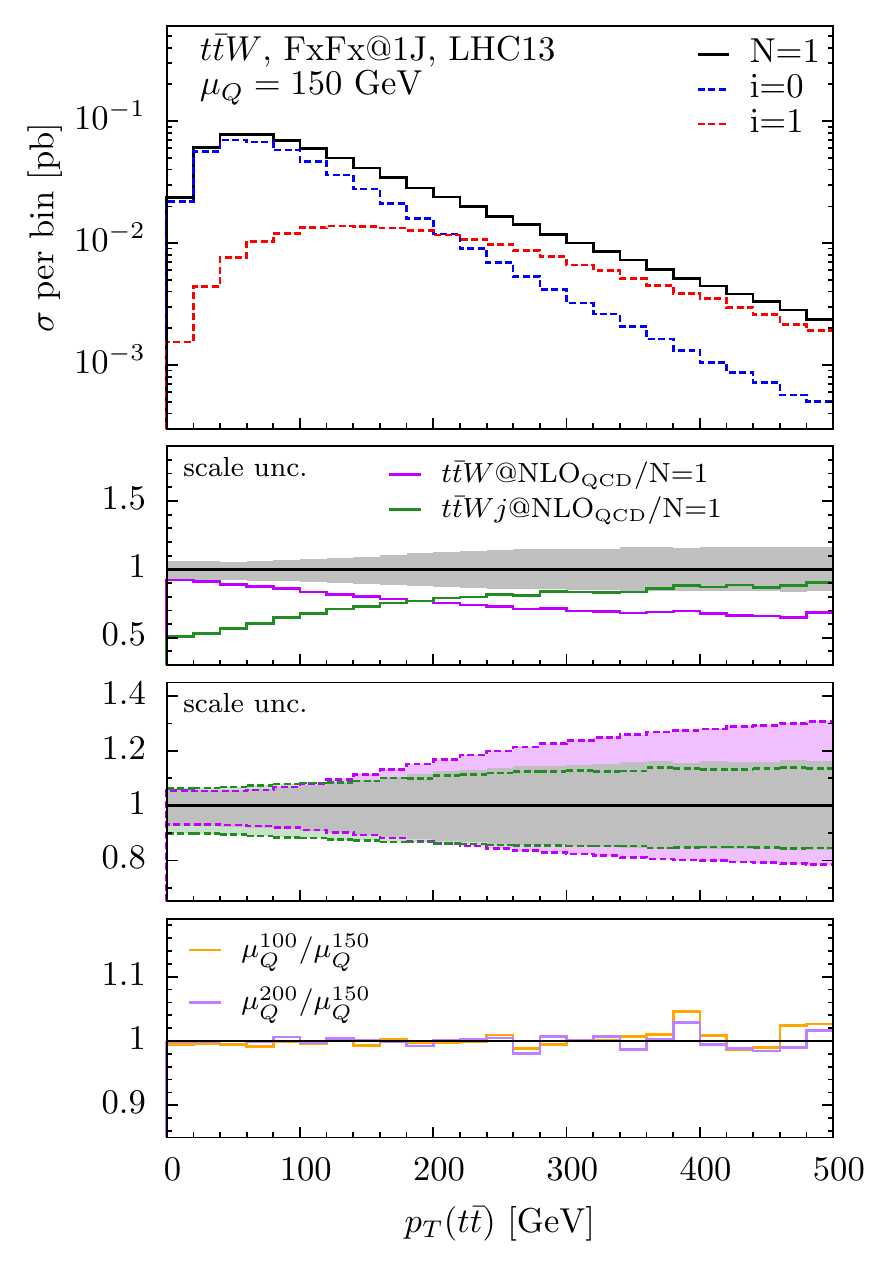}
\includegraphics[width=0.475\textwidth]{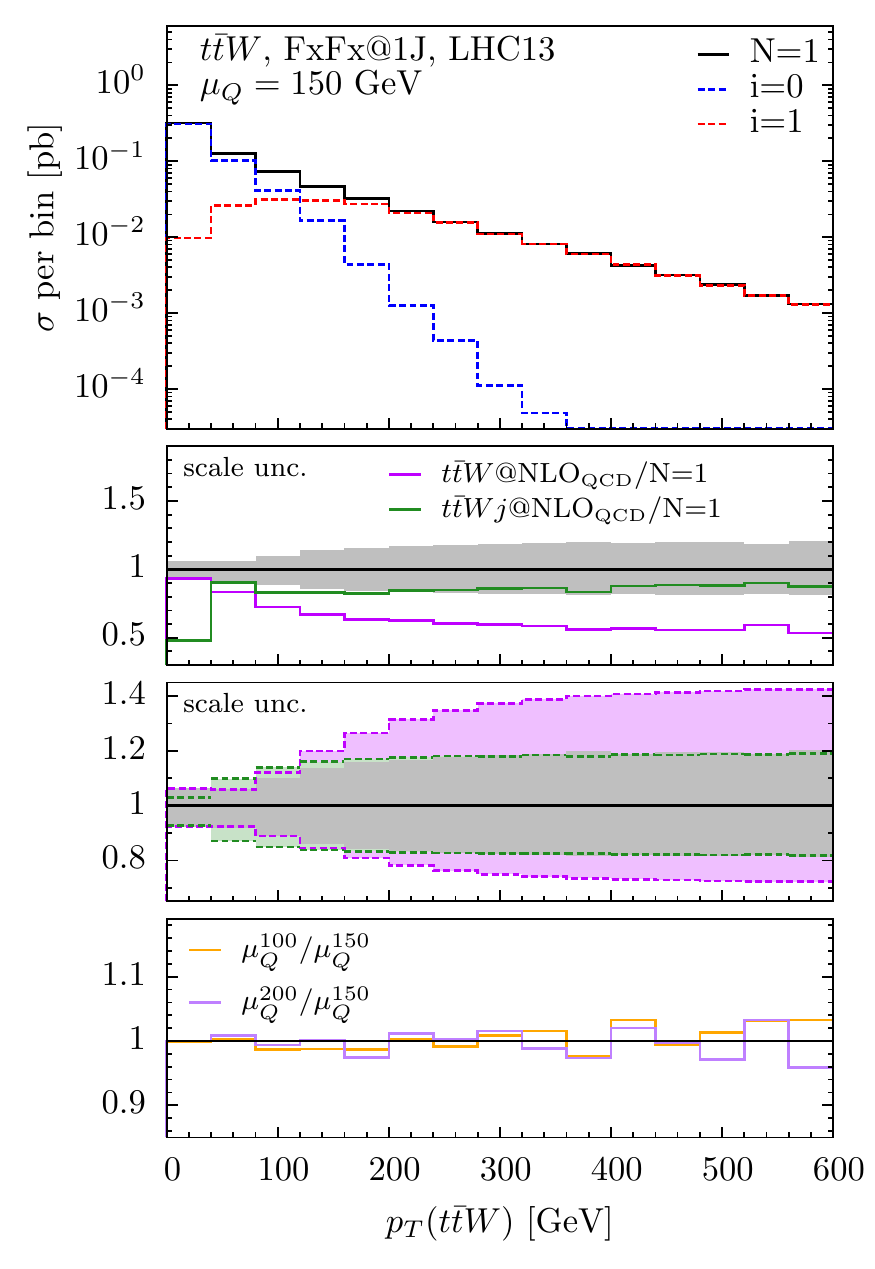}
\caption{Differential distributions for the $t \bar t W$ {\sc\small FxFx} merged sample using $\mu_Q=150$ GeV as merging scale.}
\label{fig:267_distr}
\end{figure}

In this section we investigate differential distributions of the $t$, $\bar{t}$ and $W$ at the production level. We remind to the reader that for this part of our work we keep the $t$, $\bar t$ and $W$ resonances stable, and therefore do not include hadronisation. The format of the plots in this section will be the same as the plots of Fig.~\ref{fig:267_Qcut}. In Fig.~\ref{fig:267_distr} we show some representative distributions at this level.
\noindent
In the first inset of the $p_T(t)$ and $m(t\bar t)$ distributions (upper plots) one can see that there is an overall relatively flat $K-$factor with respect to the $t \bar t W@\textrm{NLO}_\textrm{QCD}$ prediction. In the second inset we can see that the scale uncertainties of the $\textrm{FxFx@1J}$ prediction lay in between the ones of the $t \bar t W@\textrm{NLO}_\textrm{QCD}$ and $t \bar t Wj@\textrm{NLO}_\textrm{QCD}$ ones. These remarks are in accordance with the fact that these distributions are not heavily affected from the presence of the extra emissions. However, distributions like the $p_T(t\bar t)$ and $p_T(t\bar t W)$ (lower plots) are known to be sensitive to extra radiation and therefore show a behaviour similar to the $p_T(j_1)$ and $\sqrt{y_{01}}$ distributions of Fig.~\ref{fig:267_Qcut}. Of course there is no clear phase space separation of the $0-$jet and $1-$jet samples. Nevertheless, in the high-energy regime, the $\textrm{FxFx@1J}$ prediction can be only described within the scale uncertainties by the $t \bar t Wj@\textrm{NLO}_\textrm{QCD}$ one, with which it has the same level of scale uncertainties. The large and non flat $K-$factors of the merged prediction with respect to the $t \bar t W@\textrm{NLO}_\textrm{QCD}$ one show that the merging is necessary in order to describe these observables. Finally, in all these distributions, in the lowest inset, we can see that the variation of the merging scale does not change the $\textrm{FxFx@1J}$ prediction.

In order to enter the cross-section discussion in Sec.~\ref{sec:xsec} we need to clarify the level of control one acquires with the merging procedure on the scale uncertainties. In the next section we explore in detail and derive an understanding on the scale dependence of our results.

\subsection{Scale dependence}
\label{sec:scale}

\begin{figure}[h!]
\centering
\includegraphics[width=1\textwidth]{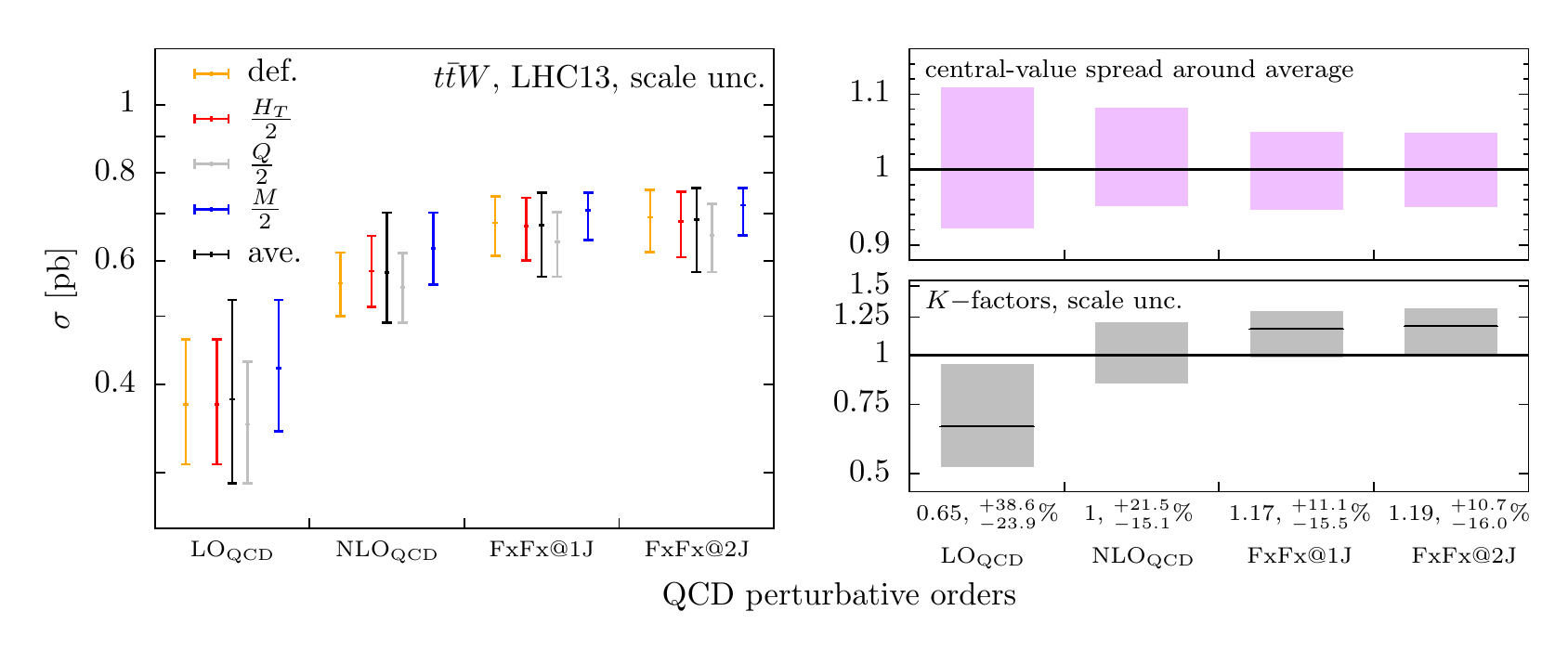}
\caption{Cross sections of various QCD perturbative orders of $t \bar t W$ production (left plot). Percent deviation of the different central values with respect to their average (upper right ratio plot). $K-$factors of the averaged central values with the combined scale uncertainties (lower right ratio plot).}
\label{fig:scale_var}
\end{figure}

Using the various functional forms for the renormalisation scale from Tab.~\ref{tab:scales} and the corresponding factorisation scale we calculate the cross section at $\textrm{LO}_\textrm{QCD}, \textrm{NLO}_\textrm{QCD}, \textrm{FxFx@1J}$ and $\textrm{FxFx@2J}$ precision. For now we omit the PDF uncertainties, since the focus of this section is on the scale dependence. In Tab.~\ref{tab:scale_var} we show the results of these different scale definitions at the various perturbative orders.
\begin{table}[h]
\renewcommand{\arraystretch}{2.0}
\begin{center}
\resizebox{\columnwidth}{!}{
\begin{tabular}{c | c c c c }
\hline
$t \bar t W$ $\sigma$[fb] & \multicolumn{4}{c}{Order} \\
\hline
Scale & $\textrm{LO}_\textrm{QCD}$ & $\textrm{NLO}_\textrm{QCD}$ & $\textrm{FxFx@1J}$ & $\textrm{FxFx@2J}$ \\
def.  & $375.1 (2)_{-67.1 (-17.9  \%)}^{+88.5 (+23.6  \%)}$ & $557.0(4)_{-57.2 (-10.3  \%)}^{+59.4 (+10.7  \%)}$ & $679.2(6)_{-69.7 (-10.3  \%)}^{+60.5 (+8.9  \%)}$ & $691.1(8)_{-74.1 (-10.7  \%)}^{+65.7 (+9.5  \%)}$ \\
$H_T/2$  & $375.1(2)_{-67.1 (-17.9  \%)}^{+88.5 (+23.6  \%)}$ & $579.5(9)_{-64.1 (-11.1  \%)}^{+70.8 (+12.2  \%)}$ & $671.9(6)_{-71.4 (-10.6  \%)}^{+64.6 (+9.6  \%)}$ & $682(1)_{-75.4 (-11.1  \%)}^{+69.1 (+10.1  \%)}$ \\
$Q/2$  & $351.0(4)_{-61.3 (-17.5  \%)}^{+80.2 (+22.9  \%)}$ & $549.7(8)_{-59.4 (-10.8  \%)}^{+65.1 (+11.8  \%)}$ & $637.7(6)_{-68.3 (-10.7  \%)}^{+65.2 (+10.2  \%)}$ & $651(1)_{-73.3 (-11.3  \%)}^{+71.5 (+11.0  \%)}$ \\
$M/2$  & $422.2(4)_{-78.8 (-18.7  \%)}^{+105.5 (+25.0  \%)}$ & $624.6(9)_{-70.2 (-11.2  \%)}^{+77.0 (+12.3  \%)}$ & $707.4(7)_{-65.7 (-9.3  \%)}^{+41.6 (+5.9  \%)}$ & $719(1)_{-68.2 (-9.5  \%)}^{+42.5 (+5.9  \%)}$ \\
ave.  & $380.8 (6)_{-91.1 (-23.9  \%)}^{+146.9 (+38.6  \%)}$  & $578(2)_{-87.4 (-15.1  \%)}^{+124.0 (+21.5 \%)}$ & $674(1)_{-104.6 (-15.5  \%)}^{+74.9 (+11.1  \%)}$ & $688(2)_{-110.1 (-16.0  \%)}^{+73.4 (+10.7  \%)}$ \\
\hline
\end{tabular}
}
\end{center}
\caption{Cross section comparisons for $t \bar t W$ using different scale definitions. The error in parenthesis is the absolute MC-error on the last digit of the central value. The scale uncertainties are shown in the form of $ \pm \{\textrm{absolute}\}(\pm \{\textrm{relative in \%}\})$.}
\label{tab:scale_var}  
\end{table}
\noindent
Furthermore we also derive the average of all the central values accompanied by the scale variation of the full envelope. In order to be easier to visualise the changes on the central values and the obtained uncertainties in each perturbative order as well as the changes from one order to the other we proceed to a pictorial representation of all the information of Tab.~\ref{tab:scale_var}. This is shown in Fig.~\ref{fig:scale_var}. In Tab.~\ref{tab:scale_var} and Fig.~\ref{fig:scale_var} one can first appreciate that after combining the different scale functional forms there is a gradual reduction of the scale uncertainty from the $\textrm{LO}_\textrm{QCD}$ to the $\textrm{FxFx@1J}$ predictions and a stability between the $\textrm{FxFx@1J}$ and $\textrm{FxFx@2J}$ ones.
\noindent
In the upper right ratio plot in Fig.~\ref{fig:scale_var} one can see the reduction of the spread of the different central scale predictions around their average by improving the accuracy of the calculation. In the lower right ratio plot along with the combined scale uncertainties we show the $K-$factors of each calculation w.r.t.~the $\textrm{NLO}_\textrm{QCD}$ prediction. One can see an extra $\sim\!\!17\%$ and $\sim\!\!19\%$ increase on the NLO QCD for the $\textrm{FxFx@1J}$ and $\textrm{FxFx@2J}$ predictions respectively. Focusing on the merged samples we point out that the scale uncertainty of the predictions with the default scale choice from Tab.~\ref{tab:scales} covers the bulk of the combined scale uncertainty. Along with all the previous remarks of this section, this is a strong indication that at this level we reach a realistic scale uncertainty (in the absence of an NNLO calculation), which is underestimated in the $\textrm{(N)LO}_\textrm{QCD}$ predictions due to the continuous opening of new channels. For this reason in the rest of the paper we will only use the default scale definition in our predictions.

\subsection{Cross section}
\label{sec:xsec}

In the previous sections we have focused on the QCD corrected perturbative orders. For any accurate cross-section prediction for $t\bar t W$ production the EW corrections must be included. It is known that they are important already at the inclusive level. It is shown in Ref.~\cite{Frederix:2017wme} that the $\sim\!\!-4\%$ of $\textrm{NLO}_\textrm{EW}^\textrm{lead}$ and the $\sim\!\!12\%$ of $\textrm{NLO}_\textrm{EW}^\textrm{sub}$ contributions with respect to the $\textrm{LO}_\textrm{QCD}$ are stable under scale variation. Furthermore, this behaviour is maintained and these numbers do not change significantly once the complete NLO corrections are further applied in the trilepton decay mode ($\sim\!\!-5.5\%$ and $\sim\!\!13\%$ respectively), after including off-shell effects and non-resonant contributions~\cite{Denner:2021hqi}. We do not include the $\mathcal O (\alpha^4)$ part of the $t\bar t W$ NLO corrections since they are at $\sim\!\!0.04\%$ level with respect to the $\textrm{LO}_\textrm{QCD}$~\cite{Frederix:2017wme} and can safely be neglected.  For our predictions in this section we also show the PDF uncertainties.

The $\textrm{NLO}_\textrm{EW}^\textrm{sub}$ contributions can be directly added within our framework since they can be matched to the parton shower~\cite{Frederix:2020jzp,Cordero:2021iau}. We separately calculate them and consistently add them to the $\textrm{FxFx@2J}$ cross section prediction. However, since the matching of the $\textrm{NLO}_\textrm{EW}^\textrm{lead}$ corrections to the parton shower is not yet done, for the inclusion of this perturbative order in the cross section prediction we calculate it at fixed order using the same settings and parameters described in Sec. \ref{sec:input_par}. The results on our cross-section predictions are presented in Tab.~\ref{tab:add_EW}. Our final prediction (last line in Tab.~\ref{tab:add_EW}) includes all the perturbative orders shown in Eq.~\ref{eq:orders}.
\begin{table}[h!]
\renewcommand{\arraystretch}{2}
\footnotesize
\begin{center}
\begin{tabular}{c | c }
\hline
Order (default scale) & $\sigma \pm \textrm{scale} \pm \textrm{PDF}$ [fb] \\
\hline
$\textrm{FxFx@2J}$ & $691.1(8)_{-74.1 (-10.7  \%)}^{+65.7 (+9.5  \%)}~_{-7.3 (-1.1  \%)}^{+7.3 (+1.1  \%)}$ \\[0.35em]
$\textrm{FxFx@2J}$+$\textrm{NLO}_\textrm{EW}^\textrm{sub}$ & $738.8(8)_{-81.3 (-11.0  \%)}^{+75.0 (+10.1  \%)}~_{-7.5 (-1.0  \%)}^{+7.5 (+1.0  \%)}$ \\
$\textrm{FxFx@2J}$+$\textrm{NLO}_{\rm EW}^{\rm lead}$+$\textrm{NLO}_\textrm{EW}^\textrm{sub}$ & $722.4(8)_{-77.7 (-10.8  \%)}^{+70.2 (+9.7  \%)}~_{-7.2 (-1.0  \%)}^{+7.2 (+1.0  \%)}$ \\
\hline
\end{tabular}
\end{center}
\caption{Addition of the EW corrections to the merged cross section. The error in parenthesis is the absolute MC-error on the last digit of the central value. The scale and PDF uncertainties are shown in the form of $ \pm \{\textrm{absolute scale}\}(\pm \{\textrm{relative scale in \%}\}) \pm \{\textrm{absolute PDF}\}(\pm \{\textrm{relative PDF in \%}\})$.}
\label{tab:add_EW}  
\end{table}
\noindent
Including all these contributions, as discussed in Sec.~\ref{sec:input_par}, there are no large missing topologies in this prediction. Furthermore, as argued in Sec.~\ref{sec:scale}, the central value is accompanied with realistic scale uncertainties, corresponding to a NLO calculation. Hence, we claim that this is currently the most-accurate estimation of the total cross section for the $pp \to t\bar{t}W$ process. The obtained cross section is increased by $\sim\!\!30\%$ w.r.t. the NLO QCD prediction (using the default scale) and is well in agreement with both the CMS and ATLAS measurements~\cite{Aaboud:2019njj,Sirunyan:2017uzs}.

\subsection{Multilepton singatures}
\label{sec:multilep}

We move in this section to the second part of our results, which are at the decay level. The details of the setup are discussed in Sec.~\ref{sec:input_par}. The $t$, $\bar t$ and  $W$ are decayed on-shell within {\sc \small MadSpin} in all possible decay modes, maintaining the tree-level spin correlations. For this level, regarding the EW corrections within our framework, following the argumentation of Sec.~\ref{sec:xsec}, we cannot include the $\textrm{NLO}_\textrm{EW}^\textrm{lead}$ contributions, but we can include the $\textrm{NLO}_\textrm{EW}^\textrm{sub}$ ones. The features of these contributions are discussed in detail at the cross-section and differential-distribution level in the multi-lepton signatures in Refs.~\cite{Frederix:2020jzp,Cordero:2021iau} and we are not going to repeat them. Our results at this level will include the perturbative orders that correspond to the second row of Tab.~\ref{tab:add_EW} and we will compare them to the NLO QCD prediction.

\begin{figure}[t!]
\centering
\includegraphics[width=0.475\textwidth]{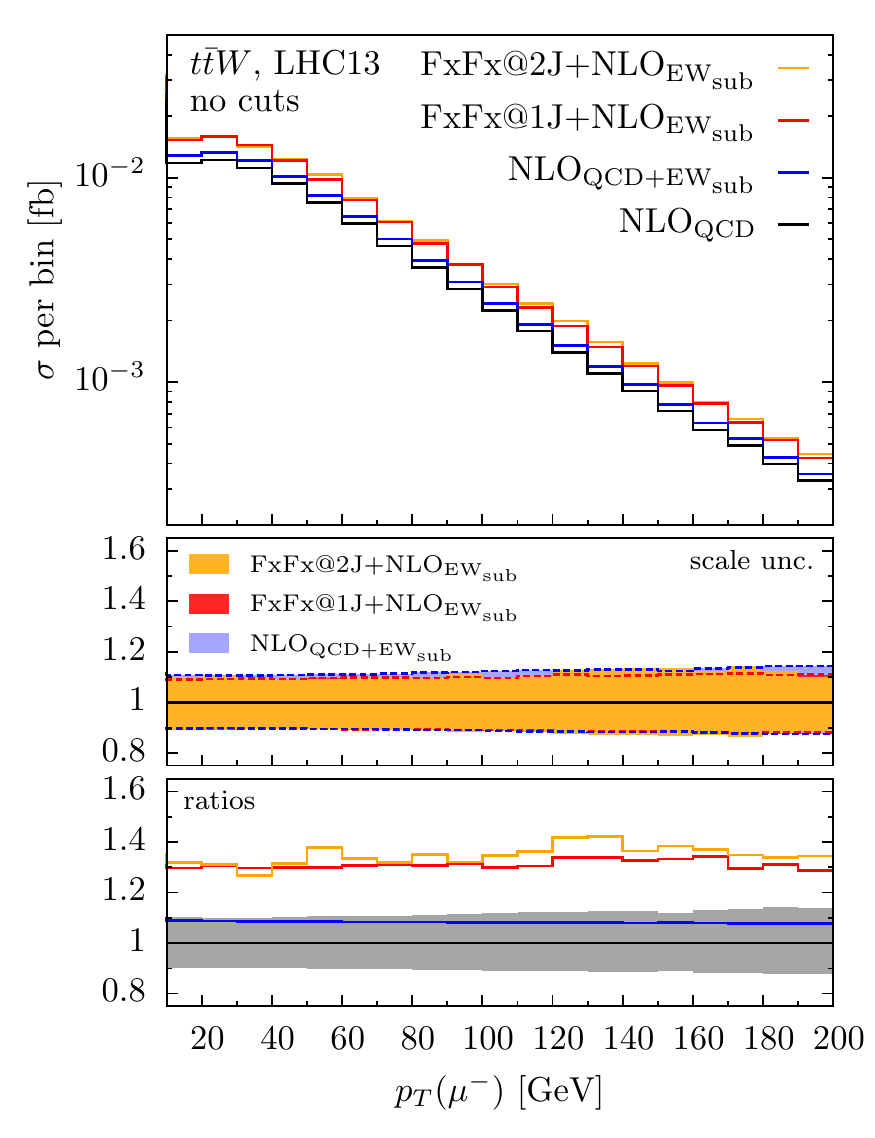}
\includegraphics[width=0.475\textwidth]{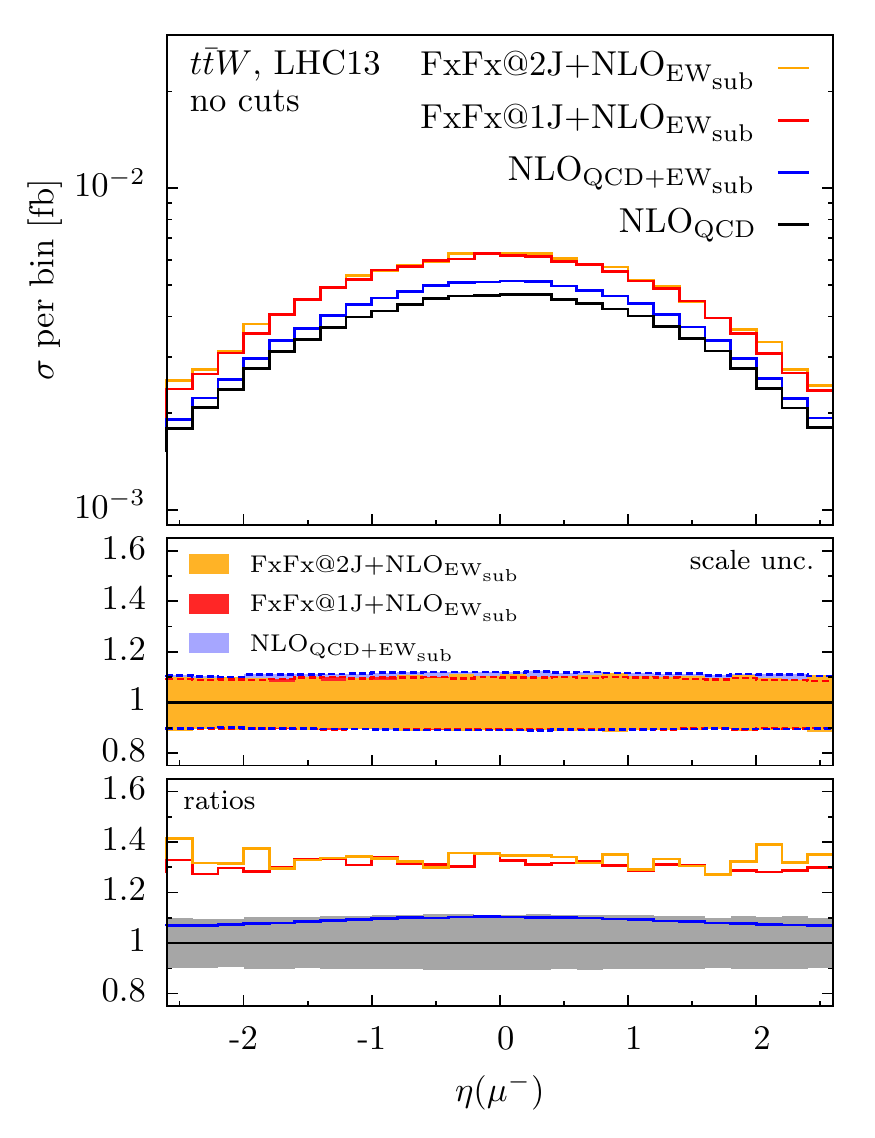} \\
\includegraphics[width=0.475\textwidth]{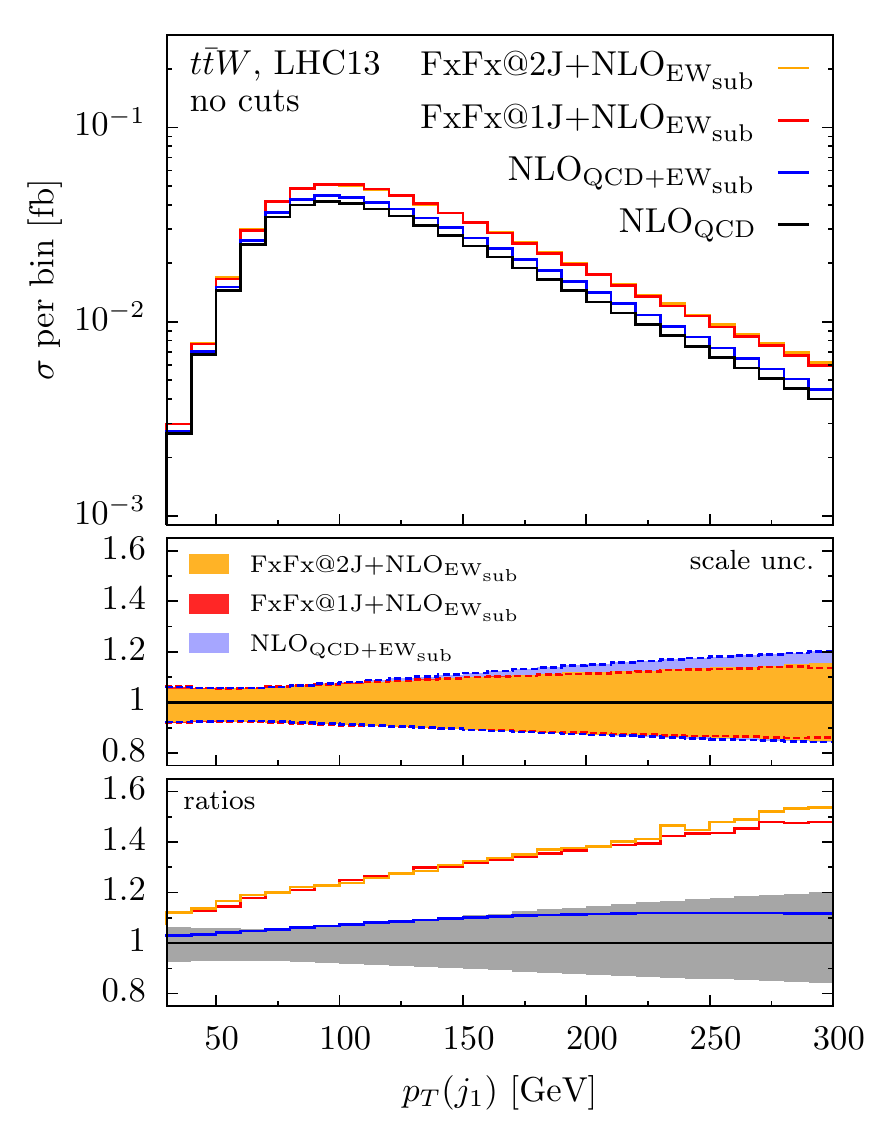}
\includegraphics[width=0.475\textwidth]{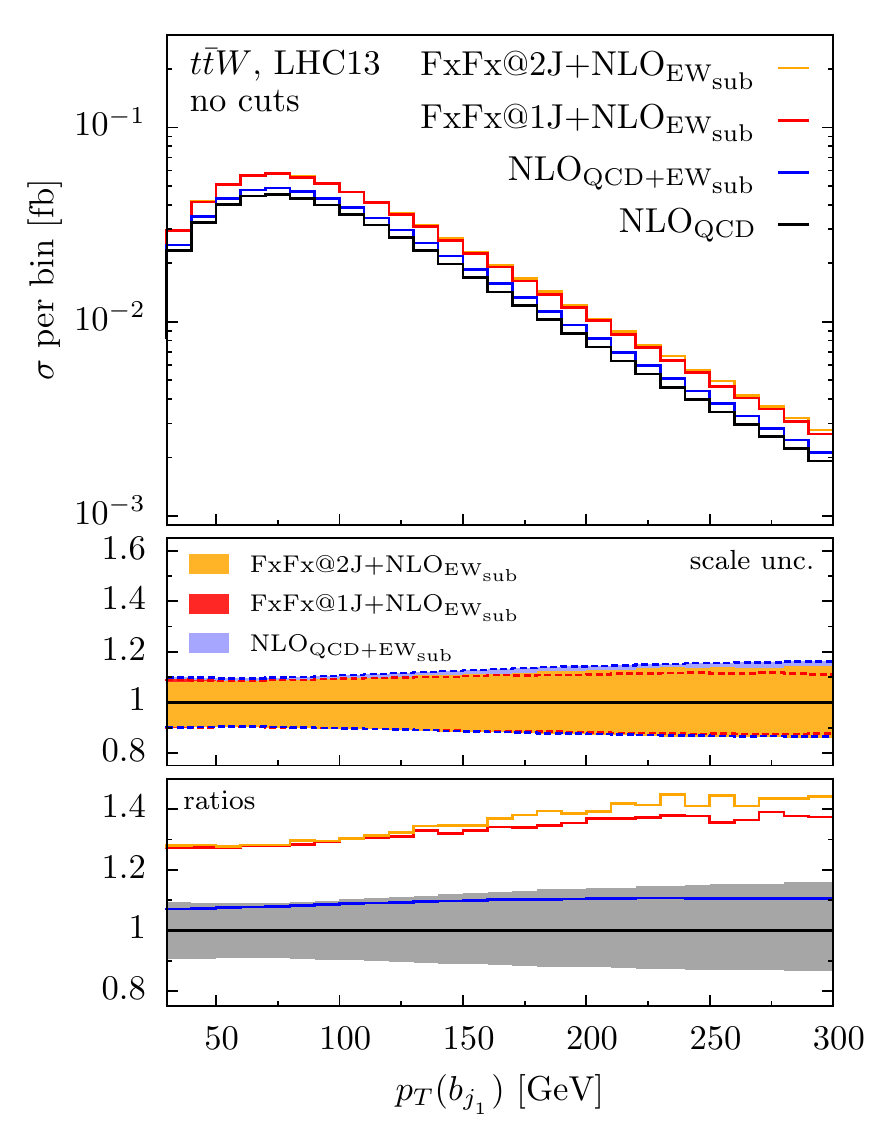}
\caption{Differential distributions at the inclusive decay level for the $t \bar t W$ process.}
\label{fig:decay_incl}
\end{figure}

Before we proceed to the specific fiducial signal regions, we start with an inclusive decay level, where we do not do apply selection or veto on the jets or leptons and we use their definitions as shown in the first two lines of Eq. \ref{eq:id}. We discuss some representative lepton and jet distributions in order to point out the relevance of our calculation. In Fig.~\ref{fig:decay_incl} we show the transverse momentum and pseudo-rapidity of the muon (upper plots) and the transverse momenta of the leading jet and $b-$jet (lower plots). Since we allow for all possible decays, the muon can emerge from a $\bar t$ or an associated $W^-$, either directly or via a leptonic $\tau^-$ decay. In the plots of Fig.~\ref{fig:decay_incl} we use the $\textrm{NLO}_\textrm{QCD}$ calculation as a reference. We then subsequently add the $\textrm{NLO}_\textrm{EW}^\textrm{sub}$ corrections and the contributions from the $1-$jet and $2-$jet merged samples.
\noindent
In the first inset we present the scale uncertainties and in the second one the ratios over the $\textrm{NLO}_\textrm{QCD}$ prediction, along with the scale variation for the latter. Regarding the $p_T(\mu^-)$ and $\eta(\mu^-)$ distributions in the last inset we can see that on top of the $\textrm{NLO}_\textrm{EW}^\textrm{sub}$ corrections the $\textrm{FxFx@1J}$ prediction adds a flat $\sim\!\!20\%$ correction. Regarding the $p_T(j_1)$ and $p_T(b_{j_1})$ distributions we see that the $\textrm{FxFx@1J}+\textrm{NLO}_\textrm{EW}^\textrm{sub}$ effect is not flat with respect to the $\textrm{NLO}_\textrm{QCD}$. In the case of the $p_T(b_{j_1})$, it grows from $\sim\!\!30\%$ to $\sim\!\!40\%$ at $300$ GeV and in the case of $p_T(j_1)$ it is significantly shaped, varying from $\sim\!\!10\%$ up to $\sim\!\!40\%$ at $300$ GeV. In the first inset we can see that in all the distributions the scale uncertainties of the $\textrm{FxFx@1J}+\textrm{NLO}_\textrm{EW}^\textrm{sub}$ are well in control and slightly reduced with respect to the $\textrm{NLO}_\textrm{QCD}+\textrm{NLO}_\textrm{EW}^\textrm{sub}$ ones\footnote{We remind to the reader that in this section, similarly to Sec.~\ref{sec:xsec} we use the default scales from Tab.~\ref{tab:scales} for all the predictions.}. In both insets we can see that the extra contributions in $\textrm{FxFx@2J}$ alter neither the scale uncertainties nor the central value of the $\textrm{FxFx@1J}+\textrm{NLO}_\textrm{EW}^\textrm{sub}$ prediction in a significant way. 

\begin{figure}[h!]
\centering
\includegraphics[width=1\textwidth]{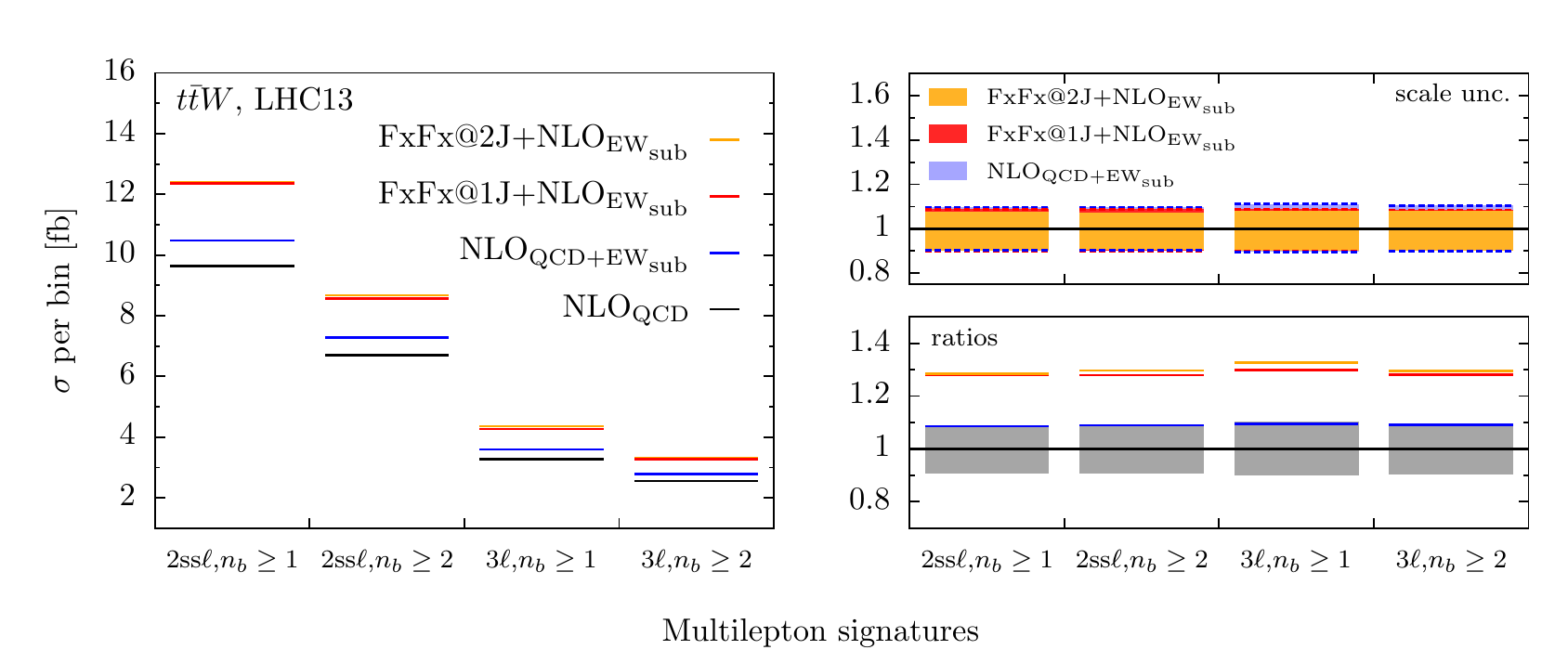}
\caption{Cross sections of various multi-lepton signal regions.}
\label{fig:fid_xsec}
\end{figure}
\noindent
\begin{figure}[h!]
\centering
\includegraphics[width=0.475\textwidth]{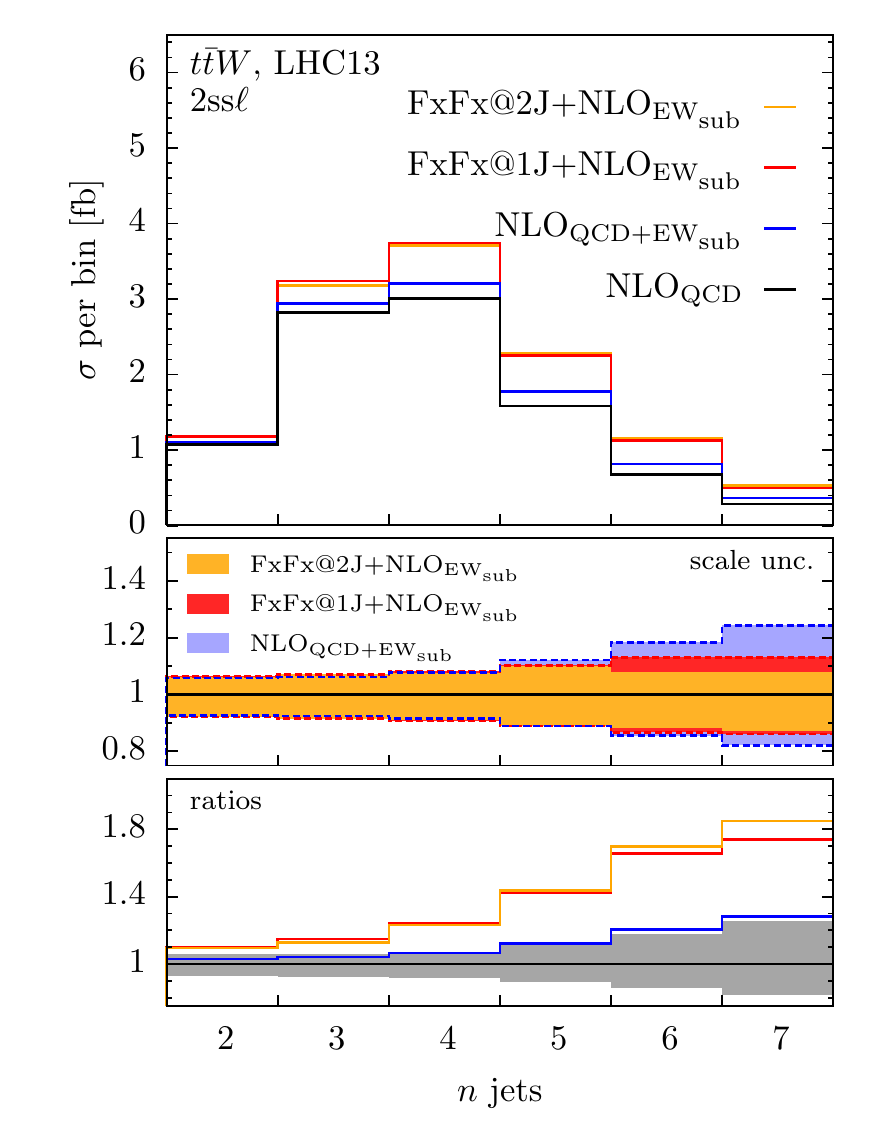}
\includegraphics[width=0.475\textwidth]{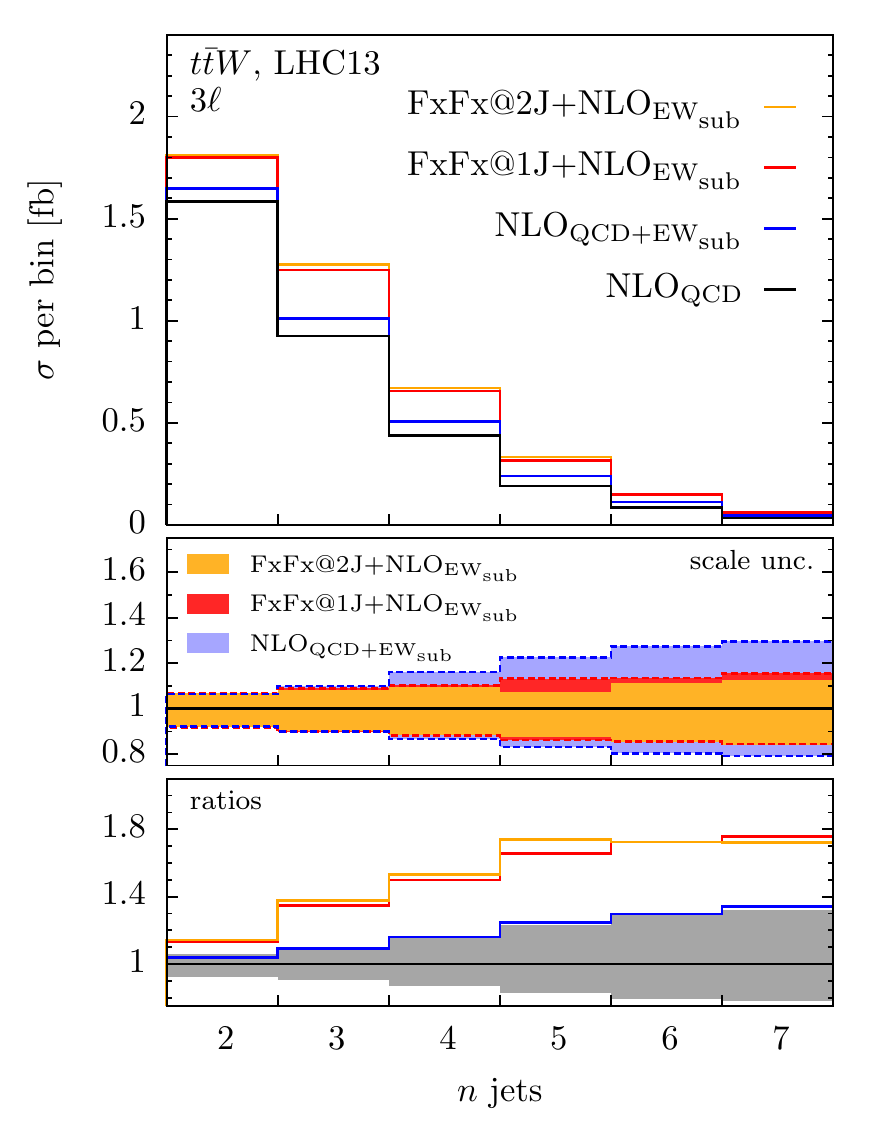}
\caption{Jet-multiplicity distributions for the $2ss\ell$ and $3\ell$ multi-lepton signatures.}
\label{fig:decay_multi}
\end{figure}
\noindent

Moving to the fiducial signatures of the multi-lepton final states, we separate the $2ss\ell$ and the $3\ell$ signal regions as defined in Eq. \ref{eq:id}. We first examine the fiducial cross section by following the same plot format, as in Fig.~\ref{fig:decay_incl}. In Fig.~\ref{fig:fid_xsec} we show the $2ss\ell$ and the $3\ell$ signal regions with at least one or two $b-$jets. In all cases we can see that the $\textrm{FxFx@1J}+\textrm{NLO}_\textrm{EW}^\textrm{sub}$ prediction induces a $\sim\!\!30\%$ increase over the $\textrm{NLO}_\textrm{QCD}$. Regarding the differential distributions in the fiducial region we focus on the jet multiplicities in both the $2ss\ell$ and $3\ell$ signatures. In Fig.~\ref{fig:decay_multi} we can see the reduction of the scale uncertainties at the tail of the distributions in the first inset. In the second inset, the $K$-factors of the $\textrm{FxFx@1J}+\textrm{NLO}_\textrm{EW}^\textrm{sub}$ prediction with respect to the $\textrm{NLO}_\textrm{QCD}$ are not flat and reach an $\sim\!\!80\%$ correction at the tails of the distributions.

\section{Conclusions and outlook}
\label{sec:concl}

In the absence of an NNLO QCD calculation for $t\bar t W$ production
and the absence of gluon induced contributions in the $t\bar t W$
soft-gluon resummation, our calculation provides a consistent way to
include the hard non-logarithmically enhanced radiation at NLO in
QCD. In this project we show the complications arising in the merging
procedure regarding the Weak-jet contributions and describe the
solution we implement in our calculation within the {\sc\small
  MadGraph5\_aMC@NLO FxFx} framework.

At the production level we show the independence of our results with
respect to the choice of the merging scale at differential and
cross-section level and further check that there is no underestimation
of the scale uncertainties. The $K$-factors with respect to the NLO
QCD prediction of differential distributions sensitive to extra
emissions (e.g.~$p_T(t\bar t), p_T(t\bar t W)$) are large and
shaped. By studying four different functional forms for the
renormalisation and factorisation scales, we demonstrate the reduction
of the scale uncertainties with respect to the NLO QCD prediction and
the significant cross-section $K$-factors. For the total inclusive
cross section we provide a prediction including the EW corrections.

At the decay level we show on the one hand that the emerged lepton
distributions follow the production-level $K$-factors but they are
flat. On the other hand the jet-related distributions (e.g.~$p_T(j_1),
p_T(b_{j_1})$) have large and non-constant $K$-factors. At the various
multi-lepton signatures there is an increase of the cross section in
agreement with the results from the production level. Finally we
present the jet-multiplicity distributions showing the shaped
$K$-factors reaching an $\sim\!\!80 \%$ correction at the tails with
respect to the NLO QCD prediction.

The predictions shown in this work are currently the most-accurate
predictions for this process, in particular at the production level,
where, due to the pecularities of the $pp\to t\bar{t}W$ process the
NLO merging improves also inclusive observables. This is because new
topologies at the NLO QCD real-emission level contribute
significantly. These, non-IR sensitive contributions can be upgraded
from tree-level to the NLO in QCD accuracy through the FxFx merging
procedure. Because the merging procedure increases the NLO QCD cross
section significantly, the observed tension between the data and the
theory is resolved.

Our implementation of the Weak-jet contributions in the FxFx merging
procedure is completely general. We will leave it for further studies
to investigate the impact of the Weak-jet contributions in other
processes.

\section*{Acknowledgments}

This work is done in the context of and supported by the Swedish
Research Council under contract number 2016-05996. IT is supported also by the MorePheno ERC grant agreement under number 668679. Computational resources to IT have been provided by the Consortium des \'Equipements de Calcul Intensif (C\'ECI), funded by the Fonds de la Recherche Scientifique de Belgique (F.R.S.-FNRS) under Grant No. 2.5020.11 and by the Walloon Region. 

\bibliographystyle{hieeetr}
\bibliography{ttW_FxFx}

\end{document}